\documentclass{aa}

\usepackage[dvips]{graphicx}
\usepackage{txfonts}
\usepackage{times,epsfig}
\usepackage{natbib} 
\bibpunct{(}{)}{;}{a}{}{,}
 
\newcommand{\Msun}{\mathrm{M}_{\odot}}
\newcommand{\Zsun}{\mathrm{Z}_{\odot}}

\newcommand{\ud}{\mathrm{d}}

\begin{document}
%
%
   \title{Genesis and evolution of dust in galaxies in the early Universe}
   
   \subtitle{I. Modelling dust evolution in starburst galaxies}
%
%
   \author{C. Gall \inst{1}, A. C. Andersen \inst{1},
          \and
          J. Hjorth \inst{1}
          }

   \institute{\inst{1}Dark Cosmology Centre, University of Copenhagen, Niels Bohr Institute, Juliane Maries Vej 30, DK-2100 Copenhagen, Denmark
                   \\
             }

   \date{Received January 12, 2011}
%
%
  \abstract
   {}
   {The aim is to elucidate the astrophysical conditions required for generating 
   large amounts of dust in massive starburst galaxies at high redshift.
   }
   {We have developed a numerical galactic chemical evolution model. 
   The model is constructed such that the effect of a wide range of parameters 
   can be investigated. 
   It takes into account results from stellar evolution models,
   a differentiation between diverse types of core collapse supernovae (CCSN),
    and the contribution of asymptotic giant branch (AGB) stars in the mass range 3--8 $\Msun$.   
   We consider the lifetime-dependent yield injection into the interstellar medium (ISM) by all sources, 
   and dust destruction due to supernova (SN) shocks in the ISM.
   We ascertain the temporal progression of the dust mass and the dust-to-gas and dust-to-metal mass ratios,  
   as well as other physical properties of a galaxy, and study their dependence on the mass of the galaxy, 
   the initial mass function (IMF), dust production efficiencies, and dust destruction in the ISM. 
   }
   {The amount of dust and the physical properties of a galaxy strongly depend on the initial gas mass available. 
    Overall, while the total amount of dust produced increases with galaxy mass, the detailed outcome
    depends on the SN dust production efficiency, the IMF, and the strength of 
    dust destruction in the ISM. 
    Dust masses are higher for IMFs biased towards higher stellar masses, 
    even though these IMFs are more strongly affected by dust destruction in the ISM. 
    The sensitivity to the IMF increases as the mass of the galaxy decreases.
    SNe are primarily responsible for a significant enrichment with dust at early epochs ($<$ 200 Myr).
    Dust production with a dominant contribution by AGB stars is found to be insufficient to account for 
    dust masses in excess of 10$^{8}$ $\Msun$ within 400 Myr after starburst.  
    }
   {We find that galaxies with initial gas masses between 1--5 $\times$ 10$^{11}$ $\Msun$ are massive enough
   to enable production of dust masses $>$ 10$^{8}$ $\Msun$. 
   Our preferred scenario is dominated by SN dust production 
   in combination with top-heavy IMFs and
   moderate dust destruction in the ISM.
   }


  \titlerunning{Genesis and evolution of dust in galaxies in the early Universe}    
  \authorrunning{C.Gall et al}  
   \maketitle
%
%
\section{Introduction}
\defcitealias{nom06}{N06}
\defcitealias{eld08}{EIT08}
\defcitealias{woos95}{WW95}

%
Modelling the evolution of dust in galaxies is a key ingredient in understanding the origin 
of the high observed dust masses in high-redshift galaxies and quasars (QSOs).
Dust masses  $\ge$ $10^{8}$ $\Msun$ have been derived from observations in QSOs at 
redshift $z$ $\gtrsim$ 6  \citep[e.g.,][]{bertol03, robs04, beel06, hin06, mich10b}
along with high star formation rates (SFR) up to a few times 10$^{2-3}$ $\Msun$ yr$^{-1}$ \citep[e.g.,][]{walt04, wan10}.
Additionally, QSOs at $z$ $>$ 6 harbour supermassive black holes (SMBHs) 
with masses $>$ 10$^{9}$ $\Msun$  \citep[e.g.,][]{will03, vest04, jiang06}. 
\cite{kaw08, kaw09} showed that to form a SMBH $>$ 10$^{9}$ $\Msun$, 
a high mass supply of  $>$ 10$^{10-11}$ $\Msun$ is needed.  
These requirements, together with derived molecular gas masses from CO line emission 
measurements in excess of 10$^{10}$ $\Msun$  \citep[e.g.,][]{cox02, caril02, bertol03b, walt03, rie09, wan10}, 
set significant constraints on the physical properties of the host galaxies. 
These in turn have implications for the origin and evolution of dust. 
Furthermore, a tendency toward increased dust attenuation with higher galaxy masses for systems 
at $z \sim 6$--8 has been found by \citet{schaer10}.

Analytical and numerical models for dust evolution have been developed.
\citet{dwe07} propose 1 $\Msun$ of dust per SN 
will be necessary to account for dust masses in high-$z$ QSOs,
 while contemplating SNe as the only source. 
Such high dust masses for SNe contradict derived dust masses from nearby SNe and SN remnants, 
which on average reveal a few times  10$^{-4}$--10$^{-2}$ $\Msun$ of dust 
\citep[e.g.,][]{woo93, elm03, tem06, mei07, rho08, kot09, sib09, barl10}. 
A review of observationally and theoretically derived dust from stellar sources 
is provided by Gall et al. (in prep, herafter GAH11).

The issue of whether SNe produce large amounts of dust is unclear. 
Other sources of dust such as AGB stars have been taken into account 
 \citep{mor03, val09} in chemical evolution models of high-redshift galaxies. 
\citet{val09} claim that with the contribution of AGB stars, 
$10^8$ $\Msun$ of dust can be reached, with AGB stars dominating the dust production. 
AGB stars (0.85--8 $\Msun$) are the main source of dust in the present universe, 
but only stars with masses $\gtrsim$ 3 $\Msun$ are likely to contribute at 
$z$ $>$ 6 \citep[e.g.,][]{marc06}.
Evidence that metal-deficient AGB stars also 
undergo strong mass loss and are able to efficiently produce dust 
is supported observationally  
\citep[e.g.,][]{zijl06, groe07, lag07b, mats07, slo09} and theoretically  \citep[e.g.,][]{wac08, mas08}. 
However, the theoretical models  \citep{dwe07, mor03, val09} greatly differ with respect to the assumptions 
made for the mass of the galaxy and dust contribution from stellar sources, as well as the treatment of the star formation.
Thus the origin of dust and its evolution remain unclear.    

In GAH11 we discuss plausible dust production efficiency limits for stellar sources between 3 $\Msun$ and 40 $\Msun$,
and determined the dust productivity of these sources for a single stellar population. 
In this paper, we investigate the evolution of dust in high-$z$ galaxies. 
We develop a numerical chemical evolution model, which allows exploration of the 
physical parameter space for galaxies at $z >$ 5--6. 
Different types of core collapse supernovae are specified and the contribution from AGB stars 
and the impact of SMBHs are taken into account. 
We furthermore follow the evolution of some physical properties of these galaxies. 
The main parameters 
varied in the model are the IMF, the mass of the galaxy, yields for SNe, and
the strength of dust destruction in the ISM, as well as the dust production efficiency limits.
Models with or without the SMBH formation are considered.  

The paper is arranged as follows.  
In Sect. \ref{SEC:MMEG} the equations used to construct the model are developed. 
We discuss the model parameters and their possible values in Sect. \ref{SEC:MOP}. 
A detailed analysis of the results is presented in Sect. \ref{SEC:RES}, 
which is followed by a discussion in Sect. \ref{SEC:DISC} 
and our conclusions of this work in Sect. \ref{SEC:CONCL}.    
%
%
%
\section{Modelling the evolution of dust in starburst galaxies}
\label{SEC:MMEG}
%
In this section we formulate the equations needed to follow a galaxy's 
time-dependent evolution in a self-consistent numerical model.
The main basic logic is adopted from \citet[][and references therein]{tins80}, which has also been 
used in other chemical evolution models to study dust in galaxies  \citep[e.g.,][]{mor03, dwe07}. 
We focus on an elaborate treatment of dust from different types of CCSNe and AGB stars. 
In order to calculate the amount of dust from these sources we use the dust production efficiencies described in Sect.~\ref{SSC:EDESNAGB}. 
In particular, the lifetime-dependent delayed dust and gas injection from AGB stars and SNe is taken into account.
The metallicity-dependent lifetimes of all stars are taken from  
\citet{schal92}, \citet{sch93}, and  \citet{cha93}. 
In GAH11 we show that the variation of the lifetime with metallicity is minimal. 
We therefore calculate and use a metallicity-averaged lifetime for all stars.
The recycled gaseous material is defined as the remaining ejected stellar yields from all 
massive stars in the mass range 3--100 $\Msun$, which has not been incorporated into dust grains. 
This also includes the stellar feedback from very massive stars ($\gtrsim$ 30--40 $\Msun$) 
in the form of stellar winds.
Dust and gas are assumed to be released instantaneously after the death of the stars.

We strictly treat the elements in the gas and solid phases separately, while paying attention to their interplay. 
Thus, we define $M_{\mathrm{g}}(t)$ as the total mass of elements in the ISM, 
which are in the gas phase and $M_{\mathrm{d}}(t)$ as the total amount of elements in the solid dust phase. 
The mass of the ISM is defined as $M_{\mathrm{ISM}}(t) \equiv M_{\mathrm{g}}(t) + M_{\mathrm{d}}(t)$, 
which in the literature is often referred to as the `gas mass'. 
The total amount of dust in our models is solely calculated from the dust contributions from 
SNe and AGB stars. 
Thus, no further growth in the ISM is contemplated. However, dust destruction in the ISM through 
SN shocks is taken into account. 
The formation of the SMBH is considered as a simple sink for dust, gas, and metals. 

We assume a so-called `closed box' model; i.e.,
the effect of infalling and outflowing gas in the galactic system is neglected. 
Although galaxies may not evolve in such a simple manner, this assumption is plausible 
since massive starburst galaxies are assumed to have SFRs $\psi(t)$ $\gtrsim$ 10$^{3}$ $\Msun$ yr$^{-1}$. 
Infall of neutral gas would only affect the system when the infall rate is comparable to the SFR. 
In this case a large gas reservoir needs to be present in the vicinity of the galaxy already. 
Besides, infall rates for high-$z$ galaxies are not known. 
We further assume the ISM to be homogeneously mixed, and we evolve our model only up to the first Gyr. 
Examples of `closed box' models being sufficiently accurate include the works of \citet{fra99} and \citet{tec04} 
for the luminous and massive submillimetre galaxy SMMJ14011+0252 at $z$ = 2.565.
%
   \subsection{Basic considerations}
   \label{SSC:BSC}
%
The initial mass functions (IMF) $\phi(m)$ is normalized to unity in the mass interval [$m_1$, $m_2$] as

\begin{equation}
\label{EQ:IMF}
         \int_{m_{1}}^{m_{2}}  m \,  \phi(m) \, \ud m = 1 ,                               
\end{equation}
where $m_1$ and $m_2$ are the lower and upper limits of the IMF (see Sect.~\ref{SSC:IMF}).

The relation between the total mass $M_{\mathrm{ISM}}(t)$ and the SFR $\psi(t)$ is given 
by the Kennicutt law \citep{kenn98}, where $\psi(t) \propto M_{\mathrm{ISM}}(t)^k$. 
Analogously to \citet{dwe07}, we apply the following notation to calculate the SFR 

\begin{equation}
\label{EQ:SFR}
         \psi(t) = \psi_{\mathrm{ini}} 
         \left[  \frac{ M_{\mathrm{ISM}}(t)}{M_{\mathrm{ini}}}\right] ^k   \, ,
\end{equation}
where $\psi_{\mathrm{ini}}$ is the initial SFR, $M_{\mathrm{ini}}$ the initial gas mass of the galaxy, 
and $M_{\mathrm{ISM}}(t)$ the mass of the ISM.  
The value of $k$ is between 1 and 2, so we set $k$ = 1.5. 
This value has often been assumed in other models \citep[e.g.,][]{dwe98, dwe07, cal08}. 
%
%
   \subsection{Equations for AGB stars and  supernovae}
   \label{SSC:ESNAGB}
%
The amount of dust from all stellar sources released into the ISM per unit time is simply calculated 
as the amount of dust produced by the stellar sources times the source rate.
The considered dust producing stellar sources are AGB stars and CCSNe. 
We account for a potentially diverse dust contribution by different SNe subtypes and
 distinguish between Type IIP SNe and the remaining Type II subtypes. 
Types Ib and Ic SNe, collectively referred to as Ib/c, are not considered as dust producing SNe. 
However they inject their stellar yields into the ISM. 
A specification of these sources and their lower and upper stellar mass limits, 
$m_{\mathrm{L(i)}}$ and  $m_{\mathrm{U(i)}}$,  are discussed in Sect.~\ref{SSC:STY}. 
In the following equations these sources are indicated by the subscript i = AGB, IIP, II, Ib/c.

The AGB and SN rate $R_{\mathrm{i}}(t)$ calculates as

\begin{equation}
\label{EQ:SNAGBRAT}
         R_{\mathrm{i}}(t) = \int_{m_{\mathrm{L(i)}}}^{m_{\mathrm{U(i)}}}
                                                \psi(t - \tau) \,
                                                 \phi(m) \, \ud m   ,  
\end{equation}
where $\tau$ = $\tau(m)$ is the lifetime of a star with a zero-age main sequence (ZAMS) mass $m$; 
i.e., the star was born at time $(t - \tau)$ when it dies at time $t$. 
It is evident that stars only contribute when the condition $t - \tau \ge 0$ is fulfilled. 

For SNe, the time of releasing the total produced elements is assumed to take place right after explosion.
The main sequence lifetime for AGB stars is defined as the time until the end of the early 
AGB phase, which can take up to several 100 Myr.
However, the most efficient mass loss phase itself is less than 1~Myr at the very end of the AGB phase 
and is relatively short compared to the total lifetime of AGB stars. 
Thus, we make the same approximation as for SNe: All produced elements and dust 
are released instantaneously after the main sequence lifetime. 
 
For each kind of source the total produced dust per unit time is calculated as
\begin{equation}
\label{EQ:SNAGBDUS}
         E^{\mathrm{prod}}_{\mathrm{d,i}}(t) = \int_{m_{\mathrm{L(i)}}}^{m_{\mathrm{U(i)}}}
                                                                        (Y_{\mathrm{Z}} +  Y_{\mathrm{WIND}}  \, Z(t - \tau)) \,
                                                                         \epsilon_{\mathrm{i}}(m,Z) \,
                                                                       \psi(t - \tau) \,
                                                                       \phi(m) \, \ud m ,   
\end{equation}
where $Y_{\mathrm{Z}}$ = $Y_{\mathrm{Z}}(m,Z)$ with $Z = Z(t - \tau)$ is the mass ($m$) 
and metallicity ($Z$) dependent amount of ejected heavy elements per star. 
The metallicity $Z$ is defined as the metallicity with which the star was born at a time $(t - \tau)$. 
The parameter $\epsilon_{\mathrm{i}}(m,Z)$ is defined as the dust production efficiency. 
The assumed efficiencies are described further in Sect.~\ref{SSC:EDESNAGB}.
For Types Ib/c SNe, $E^{\mathrm{prod}}_{\mathrm{d,Ib/c}}(t) = 0$. 
 
The total amount of mass lost in stellar winds prior to explosion, $Y_{\mathrm{WIND}}$, 
caused by mass loss during stellar evolution of SNe, is calculated as
\begin{equation}
\label{EQ:SNWIND}
         Y_{\mathrm{WIND}} = m - M_{\mathrm{fin}} ,
\end{equation}
where $Y_{\mathrm{WIND}}$ = $Y_{\mathrm{WIND}}(m,Z)$ and 
$M_{\mathrm{fin}}$ = $M_{\mathrm{fin}}(m,Z)$. 
The latter is the final mass of a SN before explosion. 
Type II SN suffer strong mass loss leading to the formation of dense circumstellar disks. 
Dust has been found in these disks for some SNe (GAH11 and  references therein).  
However,  the actual amount of metals in these disks is not known and data are not available, 
thus we cannot account for them. 
In the case of AGB stars and for SN models where no mass loss prescription is
available 
$Y_{\mathrm{WIND}}$ = 0.

Besides the elements bound in dust grains, elements in the gas phase are also released into the ISM.
The produced amount of elements in gaseous form is calculated as 
 \begin{equation}
\label{EQ:SNAGBRES}
          {E}_{\mathrm{g,i}}(t)  =  \int_{m_{\mathrm{L(i)}}}^{m_{\mathrm{U(i)}}}
                                                 (Y_{\mathrm{E}}  +Y_{\mathrm{WIND}})
                                               \psi(t - \tau) \, 
                                              \phi(m) \, \ud m  
                                              - E^{\mathrm{prod}}_{\mathrm{d,i}}(t),   
\end{equation}
where $Y_{\mathrm{E}}$ = $Y_{\mathrm{E}}(m,Z)$ is the amount of all ejected elements per star.  

The total mass of heavy elements released into the ISM per unit time is calculated as

\begin{equation}
\label{EQ:SNAGBMET}
         E_{\mathrm{Z,i}}(t) = \int_{m_{\mathrm{L(i)}}}^{m_{\mathrm{U(i)}}}
                                                    (Y_{\mathrm{Z}} + Y_{\mathrm{WIND}}  \, Z(t - \tau)) \,
                                                    \psi(t - \tau) \,
                                                    \phi(m) \, \ud m .   
\end{equation}
We include the possibility of dust destruction in the SN remnant (SNR) due to reverse shock interaction.  
Dust destruction time scales up to  $10^{4}$ yr have been predicted by  \citet{bia07} and \citet{noz07, noz10}. 
However, this timescale is relatively short in comparison to the lifetime of a star, so we make the approximation that dust is destroyed immediately after formation. 

We define the parameter $\xi_{\mathrm{SN}}$ as the SN dust destruction factor; i.e.,
the destroyed mass of dust per unit time of all SNe is 
 
 \begin{equation}
\label{EQ:SNDES}
         E^{\mathrm{dest}}_{\mathrm{d,i}}(t) = E^{\mathrm{prod}}_{\mathrm{d,i}}(t) \, \xi_{\mathrm{SN}} .
\end{equation}
This term applies only to Type IIP and Type II SNe, thus i = IIP, II.

The final SN dust injection rate per unit time is calculated as 

 \begin{equation}
\label{EQ:SNTDUS}
         E_{\mathrm{d,SN}}(t) =  \sum_{\mathrm{i = IIP,II}} E^{\mathrm{prod}}_{\mathrm{d,i}}(t) - E^{\mathrm{dest}}_{\mathrm{d,i}}(t), 
\end{equation}
while
the final AGB dust injection rate is

\begin{equation}
\label{EQ:AGBTDUS}
         E_{\mathrm{d,AGB}}(t) = E^{\mathrm{prod}}_{\mathrm{d,AGB}}(t) .
\end{equation}
\subsubsection{Recycled gaseous material} 
\label{SSS:RGM}
%
The recycled material from SNe, and AGB stars consists of all the mass of the elements not being 
incorporated into dust grains, thus the material is in the gas phase. 
Very massive stars ending as BHs may contribute with their stellar winds to the recycled material. 
We refer to stars that directly form a BH as the `remaining stars'. 
Pertaining to the short lifetime of very massive stars and the resulting short duration of the 
wind phase, we assume that all the elements lost in the wind phase are released after the death of the star. 
The total mass of the released elements per unit time of the remaining stars is

\begin{equation}
\label{EQ:REMREST}
            E_{\mathrm{g,R}}(t) = \int_{m_{\mathrm{L(R)}}}^{m_{\mathrm{U(R)}}}
                                                     X_{\mathrm{g,R}} \,
                                                     \psi(t - \tau) \,
                                                    \phi(m) \, \ud m ,
\end{equation}
where $X_{\mathrm{g,R}}$ =  $X_{\mathrm{g,R}}(m,Z)$ is the mass of elements released into the ISM per star,   
while the subscript `R'  stands for `remaining stars'.

For these remaining stars the following two scenarios are possible:  
(1) a supernova without display  \citep{eld04} occurs even if a BH is formed and elements are ejected
or (2) no SN occurs because no SN shock is launched.  
For the first case, the term $X_{\mathrm{g,R}} $ can either be substituted 
with  $(Y_{\mathrm{E}} +  Y_{\mathrm{WIND}})$,
or if  $Y_{\mathrm{E}}$ and  $Y_{\mathrm{WIND}}$ are not known, 
the common approximation of $X_{\mathrm{g,R}} = m - M_{\mathrm{rem}}$ 
can be made. 
The mass $M_{\mathrm{rem}} = M_{\mathrm{rem}}(m,Z)$ is the remnant mass of a star. 
In the second case we assume that no nucleosynthesis products will be ejected. 
A contribution to the gas household in the ISM comes solely from the stellar winds,
thus $X_{\mathrm{g,R}}$ = $Y_{\mathrm{WIND}}$. 

The mass of released metals from these stars per unit time is given by

\begin{equation}
\label{EQ:REMZ}
            E_{\mathrm{Z,R}}(t) = \int_{m_{\mathrm{L(R)}}}^{m_{\mathrm{U(R)}}} 
                                                     X_{\mathrm{Z,R}} \,
                                                     \psi(t - \tau) \, 
                                                    \phi(m) \, \ud m ,
\end{equation}
where  $X_{\mathrm{Z,R}}$ =  $X_{\mathrm{Z,R}}(m,Z)$ is the mass of the ejected heavy elements per star.
For the first case, $X_{\mathrm{Z,R}}$ = $Y_{\mathrm{Z}} +  Y_{\mathrm{WIND}}  \, Z(t - \tau)$.  
For the second, $X_{\mathrm{Z,R}} $ = $Y_{\mathrm{WIND}}  \, Z(t - \tau)$.

By extending the subscript i to i = AGB, IIP, II, Ib/c, R,  
the total returned mass of gaseous material to the ISM per unit time calculates as 

\begin{equation}
\label{EQ:REST}
          E_{\mathrm{g}}(t) =   \sum_{\mathrm{i}}{E}_{\mathrm{g,i}}(t) + 
                                               \sum_{\mathrm{i = IIP,II}} E^{\mathrm{dest}}_{\mathrm{d,i}}(t) .
\end{equation}
It is important to note  
that elements locked up in dust grains will return to the gas phase when dust destruction takes place.  
As a result, the amount of destroyed dust in SNRs must be added to the total amount of gas in 
the ISM (second term in Eq.~\ref{EQ:REST}). 
%
%
   \subsection{The evolution of dust and gas in the galaxy}
   \label{SSC:EQEDG}
%
The chemical evolution of a galaxy is mainly determined by the equations balancing the net amount of gas, 
dust, and heavy elements. 
\subsubsection{Effect of a super massive black hole}
\label{SSS:SMBH} 
%
We take the effect of a SMBH into account as an additional sink for the gas, dust, and heavy elements. 
We base our assumptions for the treatment of the BH on the theoretical work of \citet{kaw08, kaw09}. 
Super Eddington growth is required to form a SMBH. 
This necessitates a large mass supply of $\sim$ 10$^{10-11}$ $\Msun$ on a short supply timescale 
of $t_{\mathrm{SMBHsup}}$ $\sim$ 10$^{8}$ yr from the host galaxy to a massive circumnuclear disk. 
\citet{kaw09} find that the final SMBH mass is around 1--10 \% of the supply mass $M_{\mathrm{BHsup}}$. 

We do not treat the formation of the disk in detail, but assume that the supply mass needed 
is equal to the initial mass $M_{\mathrm{ini}}$ of our considered systems. 
We therefore only take the overall mass of the ISM needed to form the SMBH into account. 
A simple constant growth rate is calculated as 

\begin{equation}
\label{EQ:BHSUP}
            \Psi_{\mathrm{SMBH}} =  \frac{M_{\mathrm{SMBH}}}{t_{\mathrm{SMBH}}} ,
\end{equation}
where $M_{\mathrm{SMBH}}$ is the mass of the final SMBH, 
and $\Psi_{\mathrm{SMBH}}$ = const. for $t \le$ $t_{\mathrm{SMBH}}$, 
whereas $\Psi_{\mathrm{SMBH}}$ = 0 for $t > t_{\mathrm{SMBH}}$. 
The growth timescale to build up the SMBH is equal to the supply timescale  
$t_{\mathrm{SMBH}}$ = $t_{\mathrm{SMBHsup}}$.
The onset of the SMBH formation coincides with the onset of starburst of the whole galaxy.  
Where the SMBH is not taken into account, the growth rate is set to zero ($\Psi_{\mathrm{SMBH}} = 0$). 
\subsubsection{The amount of dust in the ISM}
\label{SSS:TDISM}
%
The evolution of the amount of dust $M_{\mathrm{d}}(t)$ in the galaxy is 
  
\begin{equation}
\label{EQ:DUST}
            \frac{\ud M_{\mathrm{d}}(t)}{\ud t} =  E_{\mathrm{d,SN}}(t) 
                                                                        +  E_{\mathrm{d,AGB}}(t) 
                                                                         -  E_{\mathrm{D}}(t) .                                                            
\end{equation}
The first and second terms are the dust injection rates from SNe and AGB stars contributing 
to the increase in the dust household in the ISM. 
The third term $E_{\mathrm{D}}(t)$ is defined as the total dust destruction rate. 
It determines the dust reduction through astration, as well as through the SMBH formation and destruction in the ISM
caused by SN shocks, if considered. 
The total dust destruction rate is calculated as

\begin{equation}
\label{EQ:DDEST}
                                    E_{\mathrm{D}}(t)  =  \eta_{\mathrm{d}}(t) \,
                                                                          ( \psi(t) +  \Psi_{\mathrm{SMBH}} 
                                                                          + {M_{\mathrm{cl}}  \, R_{\mathrm{SN}}(t)}) .
\end{equation}      
The variable $\eta_{\mathrm{d}}(t) = M_{\mathrm{d}}(t) / M_{\mathrm{ISM}}(t)$ is the fraction of dust 
in the ISM and will be referred to as the `dust-to-gas mass ratio'. 
For simplicity we make the assumption that the produced dust will be immediately mixed with the material in the ISM. 
The dust destruction in the ISM through SN shocks will be discussed in more detail in Sect.~\ref{SSC:DISM}.
\subsubsection{The amount of gas in the ISM}
%
The temporal evolution of the gas content in the galaxy is calculated as

\begin{eqnarray}
\label{EQ:GAS}
            \frac{\ud M_{\mathrm{g}}(t)}{\ud t} & = & E_{\mathrm{g}}(t)                                                                             
                                                                            + \eta_{\mathrm{d}}(t) \, {M_{\mathrm{cl}}  \, R_{\mathrm{SN}}(t)} {}  \nonumber \\
                                                               & &  {}  - (1 - \eta_{\mathrm{d}}(t)) \, (\psi(t) +  \Psi_{\mathrm{SMBH}}) .
\end{eqnarray}       
Here, $E_{\mathrm{g}}(t)$ is the recycled gaseous material (see Eq.~\ref{EQ:REST}). 
The second term is the amount of destroyed dust in gaseous form (see Sect.~\ref{SSS:TDISM}).
The third term accounts for the depletion of the gas in the ISM through incorporation into stars and the loss
to the SMBH. The fraction of gas in the ISM is expressed as $(1 - \eta_{\mathrm{d}}(t))$.
\subsubsection{Metallicity} 
%
Owing to the separation of the ISM mass into the material locked in dust and gas, respectively, 
it is necessary to formally take care of 
the transition of elements from the dust phase to the gas phase (due to destruction).  
For the evolution of  heavy elements we do not distinguish between their chemical states. 

The equation for the evolution of the total amount of  heavy elements is formulated as

\begin{equation}
\label{EQ:METZ}
            \frac{\ud M_{\mathrm{Z}}(t)}{\ud t} = E_{\mathrm{Z}}(t)
                                                                           -  \eta_{\mathrm{Z}}(t) \, (\psi(t) +  \Psi_{\mathrm{SMBH}}) ,                                                                                                                                
\end{equation}       
where

\begin{equation}
\label{EQ:METALL}
            E_{\mathrm{Z}}(t) =  \sum_{\mathrm{i}}E_{\mathrm{Z,i}}(t)                                                                      
\end{equation}       
is the total ejected mass of  heavy elements per unit time from all considered sources. 
The last term in Eq.~\ref{EQ:METZ} determines the reduction of  heavy elements due to astration and the loss to the SMBH. 

The total metallicity of the system is defined as 
$Z(t) =  \eta_{\mathrm{Z}}(t) =  {M_{\mathrm{Z}}(t) / M_{\mathrm{ISM}}(t)}$. 
The fraction of metals in the ISM which are bound in dust grains is 
calculated as  $\eta_{\mathrm{Zd}}(t) = M_{\mathrm{d}}(t) / M_{\mathrm{Z}}(t)$. 
The amount of metals in the gas phase is

\begin{equation}
\label{EQ:MMZG}
           M_{\mathrm{Z,g}}(t) = (1 - \eta_{\mathrm{zd}}(t)) \, M_{\mathrm{Z}}(t) .                                                                     
\end{equation}      
The gas phase metallicity is given as $\eta_{\mathrm{Zg}}(t) = M_{\mathrm{Z,g}}(t) / M_{\mathrm{g}}(t)$. 
%
%
%
\section{Model parameters}
\label{SEC:MOP}
%
%
In this section we describe the prime model parameters, along with the values used in this study. 
In particular, we consider the initial conditions of the galaxy, the IMF, 
stellar yields, and the destruction rates of dust in the ISM. 
These characterize the system and significantly influence the evolution of gas, 
dust, and metals. 
In addition we define some switches, which specify 
various possibilities for some model parameters.  
All parameters and their considered values, as well as the possibilities for the switches, are 
listed in Table~\ref{TAB:IPP}.
%
\subsection{Initial conditions}
\label{SSC:ICG}
%
The model is defined by the initial values of the parameters. 
It neither depends on nor is influenced by additional input from other models; 
i.e., it does not depend on a cosmological model. 
Therefore it can be applied to any galaxy within the accuracy limit of a `closed-box'  treatment.  
We are mainly interested in massive high-redshift galaxies
in which high dust masses, stellar masses, and H$_2$ masses 
have been inferred from observations. 
The parameters of our computed models are therefore tuned to such galaxies.

One of the main parameters is the baryonic initial gas mass $M_{\mathrm{ini}}$, which is 
equal to the total mass of the galaxy in baryons. 
A relation between $M_{\mathrm{ini}}$ and the mass of the dark matter halo 
$M_{\mathrm{DM}}$ hosting such systems is given through 
$M_{\mathrm{ini}} =  \Omega_{b} /  \Omega_{m} \, M_{\mathrm{DM}}$.
In this work we consider  four different massive galaxies with 
$M_{\mathrm{ini}}$ = 1.3 $\times$ 10$^{12}$ $\Msun$, 
$M_{\mathrm{ini}}$ = 5 $\times$ 10$^{11}$ $\Msun$, 
$M_{\mathrm{ini}}$ = 1 $\times$ 10$^{11}$ $\Msun$, 
and $M_{\mathrm{ini}}$ = 5 $\times$ 10$^{10}$ $\Msun$. 
 
In GAH11 we argue that the very first population of stars (so-called Pop III stars) are not likely to be the 
main sources  of high dust masses at high redshift. 
Thus we consider only the next generations of stars (Pops II or I). 
The formation of these stars takes place as soon as a critical metallicity of
 $Z_{\mathrm{cr}}$ $\sim$ 10$^{-6}$--10$^{-4}$ $\Zsun$ \citep{broloe03, schnei06, tum06} 
  is reached in the star-forming region. In this regard we assume an initial metallicity in accordance 
 with the critical metallicity of  $Z_{\mathrm{ini}}$ = $Z_{\mathrm{cr}}$ = 10$^{-6}$ $\Zsun$.  

Pertaining to the rather high derived star formation rates from observations of some high-$z$ massive 
galaxies and QSOs \citep[e.g.] [] {fra99, bertol03, rie07}, we consider an initial SFR of  
$\psi_{\mathrm{ini}}$ = 1 $\times$ $10^{3}$ $\Msun$  yr$^{-1}$. 
The evolution is determined using the Kennicutt law as described in 
Sect.~\ref{SSC:BSC}, Eq.~\ref{EQ:SFR}.  
We also consider a case of constant SFR where  
$\psi(t)$ =  $\psi_{\mathrm{ini}}$ = 1 $\times$ $10^{3}$ $\Msun$  yr$^{-1}$. 

In our model the onset of starburst is not directly connected to redshift,
so, the age of the galaxy is identical to the evolutionary time after starburst. 
For dusty galaxies seen at redshift 5--6 the earliest onset of starburst with very high SFRs can be 
considered to have taken place at $z$ $\backsimeq$ 10. 
For a $\Lambda$CDM universe with H$_{0}$ = 70 km s$^{-1}$ Mpc$^{-1}$, 
$\Omega_{\Lambda}$ = 0.73 and $\Omega_{m}$ = 0.27 and $\Omega_{b}$ = 0.04 \citep{sper03}, 
the evolutionary time of interest for building up high dust masses possibly lies then within 400--500 Myr. 
We have computed all models presented in this paper up to an age of the galaxy of $t_{\mathrm{max}}$ = 1 Gyr. 
 %
\subsection{The initial mass function}
\label{SSC:IMF}
%
\begin{table}
\caption{{IMF and the adopted values}}              
\label{TAB:IMF}
\centering
\begin{tabular}{lcccl}
\hline
\hline
   IMF       &$\alpha$ & $m_ {1}$ & $m_{2}$ & $m_{\mathrm{ch}}$\\                 
\hline
\\   Salpeter         &   2.35       &     0.1        &     100      & ---\\ 
     Mass-heavy &     2.35       &     1.0        &     100      & ---\\  
     Top-heavy     &      1.5        &     0.1        &     100      & ---\\
     Larson  1        &     1.35       &     0.1        &     100      & 0.35\\
     Larson  2        &    1.35       &     0.1        &     100      & 10.0\\              
\hline
\end{tabular}\\
\end{table}
%
The initial mass function is 
an important parameter influencing the evolution of dust, gas, and metals in a galaxy. 
It determines the mass distribution of a population of stars with a certain ZAMS mass.  
In models of dust evolution in galaxies and high-$z$ quasars  
\citep [e.g.,] [] {mor03, dwe07}, an IMF  first proposed by \citet{salp55} is often used.
However, for starburst galaxies there is also observational evidence of a top-heavy 
IMF \citep[e.g.][]{doa93, tum06, dab09, hab10}. 

In view of this, we consider a set of five different IMFs. 
The power-law IMFs (Salpeter, mass heavy and top-heavy) have the form 
$\phi(m)  \propto  \, m^{-\alpha}$. The lognormal Larson IMFs \citep{lars98} are given as
$\phi(m)  \propto  \, m^{-(\alpha + 1)} \exp (- m_{\mathrm{ch}} / m)$, 
where $m_{\mathrm{ch}}$ is the characteristic mass. 
Their parameters are presented in Table~\ref{TAB:IMF}.

The influence of the diverse IMFs, particularly on the dust production rates from stellar sources, is
demonstrated in GAH11. 
%
%
\begin{table*}
\caption{{List of all model parameters}}          
\label{TAB:IPP}
\centering
\begin{tabular}{llll}
\hline
\hline
   Parameters			& Value									& Unit				&Description\\                 
\hline
\\  $M_{\mathrm{ini}}$	& 5 $\times$  $10^{10}$, 1 $\times$  $10^{11}$, 
                                                  5 $\times$  $10^{11}$, 
                                                  1.3 $\times$  $10^{12}$	& $\Msun$								&Initial mass of the galaxy \\ 
     $\psi_{\mathrm{ini}}$    & 1 $\times$ $10^3$							& $\Msun$  yr$^{-1}$	& Star formation rate\\  
    $Z_{\mathrm{ini}}$         & 10$^{-6}$								&$\Zsun$				& Initial metallicity\\
    $k$				& 1.5										&					& Power for the relation $\psi(t) \propto M_{\mathrm{ISM}}(t)^k$   \\  
    $M_{\mathrm{cl}}$		& 800, 100, 0								& $\Msun$			& Swept-up ISM mass per SN\\
    $M^{\mathrm{crit}}_{\mathrm{core}}$	& 15							& $\Msun$			& Critical He core mass\\
    $\xi_{\mathrm{SN}}$	& 0.93									& 					& SN dust destruction factor\\ 
   $M_{\mathrm{SMBH}} $	& 3 $\times$  $10^{9}$, 5 $\times$  $10^{9}$		&  $\Msun$			& Mass of the SMBH\\
  $t_{\mathrm{SMBH}}$  	& 4 $\times$ $10^{8}$						& yr					& Growth timscale for the SMBH\\
  $t_{\mathrm{max}}$	& $10^{9}$								& yr					& Maximum computed age of the galaxy \\
    \\             
\hline
   Parameters			& Switch									&  					&Description\\                 
\hline
\\  $Y_{\mathrm{Z}}$, $Y_{\mathrm{E}}$,  $Y_{\mathrm{Q}}$ (for SN)
					&  \multicolumn{2}{l}{\citetalias{eld08}, 
    					 \citetalias{woos95},  \citetalias{nom06}, 
					 \citet{geo09}}													& Possibilities for the SN yields\\
    $Y_{\mathrm{Z}}$, $Y_{\mathrm{E}}$, $Y_{\mathrm{Q}}$ (for AGB)	
    					&  \multicolumn{2}{l}{\cite{vhoek97}}		 							& Possibilities for the AGB yields\\
    $\phi(m)$			&  \multicolumn{2}{l}{Salpeter, mass-heavy, 
    						top-heavy, Larson 1, Larson 2}									& Initial mass function\\
    SFR				&  \multicolumn{2}{l}{evolving / constant}								& Additional switch for the SFR\\
 $\epsilon_{\mathrm{AGB}}(m,Z)$
 					&	only one case considered		 			&					& Dust formation efficiency, AGB \\
 $\epsilon_{\mathrm{SN}}(m)$
 					& max / low       								&					& SN dust formation efficiency \\
 $\xi_{\mathrm{SN}}$  	& considered / not considered					&					& SN dust destruction \\
    BH / SN				& SN when BH / no SN when BH				&					& Possibility, if a SN occurs even a BH  \\
   		                            &  		     								&					& is formed or not\\
   SMBH				& considered / not considered					&					& Growth of SMBH \\
\hline
\end{tabular}\\
\end{table*}
%
\subsection{Stellar yields}
\label{SSC:STY}
%
The model is adapted to published results from  stellar evolution models. 
Using these models the metallicity dependent upper and lower mass limits of the mass range 
of diverse SNe subtypes can be determined.
For comparison to the usually adopted treatment of SNe we calculate simplified models with 
a fixed mass range for CCSNe between 8--40 $\Msun$ and use stellar yields from nucleosynthesis 
calculations for SNe.   
For SNe we adopt three different structural models.
\subsubsection{SN yields from stellar evolution models} 
%
We adopt the stellar yields for the hydrogen mass $M_{\mathrm{H}}(m,Z)$ in the envelope, 
the He core mass $M_{\mathrm{core}}(m,Z)$, and the final mass $M_{\mathrm{fin}}(m,Z)$ of the 
progenitors, as well as $Y_{\mathrm{E}}(m,Z)$ and $Y_{\mathrm{Z}}(m,Z)$ from the single 
stellar evolution models by \citet[][herafter  \citetalias{eld08}]{eld08}. 
These yields are based on previous works \citep{eld04, eld05}.  
\citetalias{eld08} adopt a mass loss prescription using the rates of  \citet{jag88} with the rates of  
\citet{vink01} for pre-Wolf-Rayet (WR) and from  \citet{nug00} for WR evolution including overshooting. 
In their study, this model is closest to a set of SN progenitor observations. 

\citet{eld04, eld05} also show that  all stars end up as II-P SN at metallicities roughly below $Z$ = 10$^{-4}$, 
while with increasing metallicity 
the upper mass limit for II-Ps decreases; i.e., at solar metallicity it is at roughly 28 $\Msun$. 
Type II-P SNe are per definition all stars that have retained at least 2 $\Msun$ of hydrogen in their envelopes 
at their pre SN stage \citep{heg03}. 
Since mass loss is more efficient at higher metallicities, the upper mass limit for II-P supernovae decreases. 

With increasing metallicity further types of CCSNe such as IIL, Ib, and Ic SNe arise. 
The upper limit for IILs is defined by the small hydrogen fraction of $\sim$ 0.1 $\Msun$ in the envelope. 
In case little or no hydrogen in the envelope of the progenitor is present, SNe appear as Types Ib or Ic.  
The SNe arising from the higher mass end of all Type IIs are the subtypes IIb or IIn. 
These types are difficult to fit into a quiescent mass loss prescription and are not specified by  \citetalias{eld08}. 
Thus we simply assume that these subtypes may be part of the fraction of SNe, which is determined as 
 IIL in the models of \citet{eld04}. 
These SNe will be collectively referred to as the remaining Type II supernovae. 

We define the conditions for determining the lower $m_{\mathrm{SL,(i)}}$ and upper $m_{\mathrm{SU,(i)}}$ 
boundaries of the different SN subtypes analogous to \citet{eld04} as given in Table~\ref{TAB:DSNT}.  %
\begin{table}
\caption{{Definitions of stellar types}}              
\label{TAB:DSNT}
\centering
\begin{tabular}{lcc}
\hline
\hline
   Type       		&$M_{\mathrm{H}}(m,Z)$ [$\Msun$]	&$M_{\mathrm{core}}(m,Z) /  M^{\mathrm{crit}}_{\mathrm{core}}$\\                 
\hline
\\   Type II-P		& $\ge 2$						&$\le$ 1\\ 
Remaining Type II	&$\ge 0.1$ and $< 2$			&$\le$ 1\\  
Type Ib/c			&$< 0.1$						&$\le$ 1\\
Remaining stars	&---							&$>$ 1\\
\hline
\end{tabular}\\
\end{table}
 %
The masses $M_{\mathrm{H}}(m,Z)$ and $M_{\mathrm{core}}(m,Z)$ are dependent on metallicity. 
Therefore the lower and upper mass limits for these SN types are  
$m_{\mathrm{SL,(i)}}$ = $m_{\mathrm{SL,(i)}}(Z)$ and $m_{\mathrm{SU,(i)}}$ = $m_{\mathrm{SU,(i)}}(Z)$. 
The lower mass limit for Type II-P SNe is fixed at $m_{\mathrm{SL,IIP}}(Z) \equiv  m_{\mathrm{SL,IIP}}$ = 8 $\Msun$.  
The absolute upper mass limit for the most massive stars is the cutoff mass defined by the IMF, thus
$m_{\mathrm{SU,(i)}}(Z) \equiv m_{\mathrm{SU,(i)}}$ = $m_{2}$.

The criterion for direct BH formation is based on the system used by \citet{heg03} and \citet{eld04}.
A direct BH forms when the final He core mass $M_{\mathrm{core}}(m,Z)$ 
exceeds the critical He core mass, $M^{\mathrm{crit}}_{\mathrm{core}}$ = 15 $\Msun$. 
Models using yields from \citetalias{eld08} will be referred to as `EIT08M'. 

\subsubsection{SN yields from rotating stellar models} 
%
We take the yields for $M_{\mathrm{H}}(m,Z)$, $M_{\mathrm{fin}}(m,Z)$, $Y_{\mathrm{E}}(m,Z)$, 
and $Y_{\mathrm{Z}}(m,Z)$ from \citet{geo09}.  
Rotationally enhanced mass loss becomes very efficient with increasing metallicity. 
The mass loss description used in their models are from \citet{meyma03, meyma05} 
with a rotational velocity of 300 km/s.
Provided that a SN occurs even if a BH is formed, the resulting range for Type II SN 
(without further subdivision) is between 8--54 $\Msun$ at a metallicity $Z$ = 0.004 $\Zsun$. 
However, only stars between 8--25 $\Msun$ form neutron stars. 
At $\Zsun$ the Type II mass range becomes narrower (8--25 $\Msun$) and stars above the upper limit will 
become WR stars and explode as Ib/c SNe. The range for neutron stars at this metallicity extends up to 35 $\Msun$. 

To determine the lower $m_{\mathrm{SL,(i)}}(Z)$ and upper $m_{\mathrm{SU,(i)}}(Z)$ mass limits 
for the considered SNe types and remaining stars (i = IIP, II, Ib/c, R), we appled the same definitions 
as described for the EIT08M models. 
However, the direct BH cut is constrained by the mass of the remnant star $M_{\mathrm{rem}}(m,Z)$ 
instead of $M_{\mathrm{core}}(m,Z)$.  
We follow the notation for the BH formation by \citet{geo09}, where a BH forms when 
$M_{\mathrm{rem}}(m,Z)$ $>$ 2.7 $\Msun$. 
Models where these yields and the above SN type division and BH formation 
are applied, are referred to as `G09M'.
\subsubsection{Models with fixed SN mass range} 
%
A treatment in which
the boundaries of the mass range for all considered dust-forming supernovae and remaining stars 
are fixed throughout the evolution
is quite common and has been used in previous models  
\citep[e.g.] [] {dwe98, mor03, dwe07, val09}. 

We used the stellar yields for $Y_{\mathrm{E}}(m,Z)$, $Y_{\mathrm{Z}}(m,Z)$ and certain elements 
$Y_{\mathrm{Q}}(m,Z)$ with Q = C,O  from nucleosynthesis models of either  
\citet[][hereafter  \citetalias{woos95}]{woos95}  or \citet{nom06} (herafter  \citetalias{nom06}). 
All SNe are of Type II and considered to be within the mass interval 
$m_{\mathrm{L,(II)}}$ = 8 $\Msun$ and $m_{\mathrm{U,(II)}} = 40$. 
Yields for the final mass prior to explosion are not available, thus in all equations $Y_{\mathrm{WIND}} = 0$. 

The remaining very massive stars between $m_{\mathrm{L,(R)}}$ = $m_{\mathrm{U,(II)}}$ = 40  
and the upper limit $m_{\mathrm{U,(R)}}$ = $m_{\mathrm{2}}$  are assumed to turn into BHs. 
We apply the yields from \citetalias{eld08} to these remaining stars in order to account for the gas return 
into the ISM from their mass loss phase according to Eqs.~\ref{EQ:REMREST} and~\ref{EQ:REMZ}. 
Thus, the option of whether a SN explosion occurs or not is retained. 

The models with stellar yields from \citetalias{woos95} will be referred to as `WW95M' and the 
models with yields from  \citetalias{nom06} will be referred to as `N06M'. 
\subsubsection{AGB stars} 
%
For high-mass AGB stars we use the stellar yields for the total ejected mass 
$Y_{\mathrm{E}}(m,Z)$, the metals $Y_{\mathrm{Z}}(m,Z)$ and certain elements 
$Y_{\mathrm{Q}}(m,Z)$ with Q = C,O from \citet{vhoek97}. 
The mass interval is set to a fixed mass range of $m_{\mathrm{L,(i)}}$ = 3 $\Msun$ 
and $m_{\mathrm{U,(i)}}$ = 8 $\Msun$, i = AGB. 
AGB stars with masses below 3 $\Msun$ are not considered. 
%
%
%
\subsection{AGB star and SN dust production efficiencies}
 \label{SSC:EDESNAGB}
%
%
  \begin{figure}
    \resizebox{\hsize}{!}{ \includegraphics{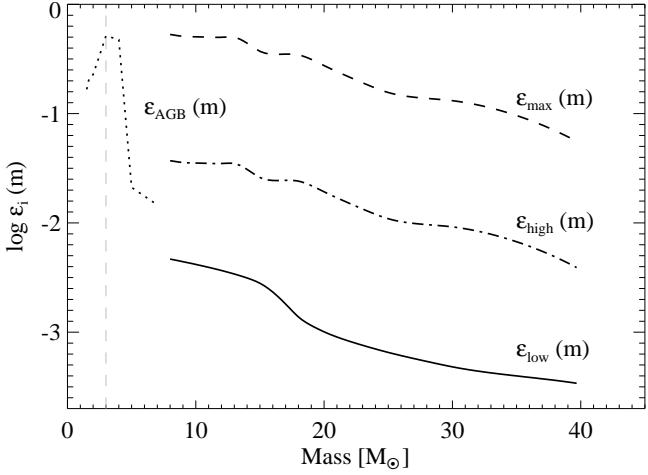}}   
    \caption{
Dust-production efficiencies of massive stars. 
The dashed line represents the `maximum' SN efficiency $\epsilon_{\mathrm{max}}(m)$ 
and the dashed dotted line the `high' SN efficiency $\epsilon_{\mathrm{high}}(m)$. 
The solid line is the `low' SN efficiency $\epsilon_{\mathrm{low}}(m)$.
The left dotted line represents the metallicity-averaged AGB efficiency 
$\epsilon_{\mathrm{AGB}}(m)$.  
The vertical line marks the least massive (3 $\Msun$) AGB star considered.
}
    \label{FIG:COFPLOT}              
   \end{figure}
The values for the AGB star and SN dust production efficiencies are
derived from observationally and theoretically determined dust yields. 
The dust production efficiency per stellar mass and 
metallicity, $\epsilon_{\mathrm{i}}(m,Z)$,  is calculated as 

   \begin{equation}
      \epsilon_{\mathrm{i}}(m,Z)  = \frac{Y_{\mathrm{d}}(m, Z)}{Y_{\mathrm{Z}}(m, Z)},
   \end{equation}
where $Y_{\mathrm{d}}(m, Z)$ is assumed to be the final amount of dust ejected into the ISM 
(i.e., produced and possibly processed through shock interactions).

For AGB stars, we adopted for $Y_{\mathrm{d}}(m, Z)$ the theoretically derived
total dust yields from  \citet{ferra06} and for $Y_{\mathrm{Z}}(m,Z)$ the yields from  \citet{vhoek97}. We calculated the AGB star efficiency for different metallicities. 
For illustration, we show the metallicity-averaged efficiency, $\epsilon_{\mathrm{AGB}}(m)$, which is representative of the trend of the metallicity-dependent $\epsilon_{\mathrm{AGB}}(m,Z)$ in Fig.~\ref{FIG:COFPLOT}. 
One notices that, on average, AGB stars between 3--4 $\Msun$ are the most efficient dust producers. 

For the dust production efficiency for SNe we used two plausible limits: 
(i) an upper limit that was motivated by theoretical SN dust formation models, 
and (ii) a lower limit based on dust masses inferred from observations of older SN remnants, such 
as Cas A, B0540$-$69.3, Crab nebula, and 1E0102.2$-$7219
 \citep [e.g.,] [] {dou01a, bou04, gree04, kra04, tem06, rho08, san08, wilb08, dun09, barl10}.  
For the upper limit we applied two different combinations of the dust and metal yields, which are 
(i)  $Y_{\mathrm{d}}(m, Z)$ from \citet{tod01} 
with $Y_{\mathrm{Z}}(m,Z)$ from \citetalias{woos95}  and 
(ii) $Y_{\mathrm{d}}(m, Z)$ from  \citet{noz03}  
with  $Y_{\mathrm{Z}}(m,Z)$  from  \citetalias{nom06}. 
Averaging the obtained efficiencies over $Z$ leads to the metallicity independent `maximum' SN dust production efficiency, 
$\epsilon_{\mathrm{max}}(m)$.  
For the lower limit an average dust mass of 3 $\times$ $10^{-3}$ $\Msun$  
inferred from the SN remnants is applied to all SNe in the mass range 8--40 $\Msun$.  
To calculate the SN dust production efficiency, stellar yields from 
\citetalias{woos95},  \citetalias{nom06}, and \citetalias{eld08} are used. 
The efficiencies obtained with these yields are averaged, constituting the `low' SN dust production efficiency, $\epsilon_{\mathrm{low}}(m)$.  

Applying a dust destruction factor (see Sect.~\ref{SSC:ESNAGB}) to  
$\epsilon_{\mathrm{max}}(m)$ results in a reduced dust production efficiency.
We refer to this efficiency as `high' SN dust-production efficiency, 
$\epsilon_{\mathrm{high}}(m)$ =  $\xi_{\mathrm{SN}}$ $\epsilon_{\mathrm{max}}(m)$. 
The amount of dust ($\sim$ 2--6 $\times$ 10$^{-2}$ $\Msun$) produced per SN
with this efficiency is comparable to the higher dust masses derived in older SN remnants.

We used a dust destruction coefficient, $\xi_{\mathrm{SN}}$ = 0.93, following \citet{bia07}.
These SN efficiencies may possibly also arise from different scenarios of dust formation, such as grain growth and shock interactions in SN remnants. 
The averaged SN  efficiencies $\epsilon_{\mathrm{max}}(m)$ and $\epsilon_{\mathrm{low}}(m)$, as well as the `high' SN efficiency, 
$\epsilon_{\mathrm{high}}(m)$, are seen in Fig.~\ref{FIG:COFPLOT}.
%
%
\subsection{Dust destruction in the ISM}
\label{SSC:DISM}
Once a dust grain is injected into the interstellar medium, it is subject to either growth or to 
disruptive or destructive processes. 
We here focus on the destructive and disruptive ones due to supernova shocks. 

Disruptive processes are those that lead to fragmentation of large dust grains 
(radius $> 1000~\AA$) into smaller dust grains (radius $< 500~\AA$). 
\citet{jon96} found that shattering due to grain-grain collisions dominates vaporization 
and therefore also determines the grain size redistribution, which is shifted towards smaller grains. 

Destructive processes return dust grains back into the gas phase.
The main destruction of dust grains in the ISM comes from the sputtering caused by interstellar shock waves 
with shock velocities $\ge$ 100 km $\mathrm {s^{-1}}$ \citep{seab87, jon96}. 
The destruction takes place because of high-velocity gas-grain impacts of smaller projectiles 
(radius $< 100~\AA$), such as energetic He$^{+}$ ions, onto dust grains. 
This results in the removal of dust species at or near the surface of the grains.

A dust grain is exposed to thermal and nonthermal sputtering, 
as well as to vaporization during the passage of the shock.
\citet{jon96} also found that graphite grains are mainly destroyed by thermal sputtering. 
Silicates are equally affected by thermal and non-thermal sputtering in 
high-velocity shocks with velocities v$_{s} > $150 km $ \mathrm {s^{-1}}$, 
while vaporization is negligible. 
These processes determine the lifetime of dust grains in the ISM. 
The timescale $\tau_{\mathrm{dl}}(t)$ of the dust grains against 
destruction, or simply the lifetime of the dust grains, 
and the dust destruction rate in the ISM $E_{\mathrm{D,ISM}}(t)$ 
are given \citep{mct89, dwe98, dwe07} through
\begin{eqnarray}
\label{EQ:DLT}
           \tau_{\mathrm{dl}}(t) & = & \frac{M_{\mathrm{ISM}}(t)}
                                                           {M_{\mathrm{cl}}  \, R_{\mathrm{SN}}(t)}  {}; \nonumber\\        
       E_{\mathrm{D,ISM}}(t) & = & \frac{M_{\mathrm{d}}(t)}{\tau_{\mathrm{dl}}(t)} ,
\end{eqnarray}       
where $M_{\mathrm{ISM}}(t)$ is the mass of the ISM and $R_{\mathrm{SN}}(t)$ 
the supernova rate of all supernovae causing the destruction. 
The mass $M_{\mathrm{cl}}$ is the mass of the ISM, which is completely cleared of dust 
through one single supernova remnant. 
These two equations combined lead to an expression for $E_{\mathrm{D,ISM}}(t)$ in the form   

\begin{equation}
\label{EQ:DESTR}
          E_{\mathrm{D,ISM}}(t) = \eta_{\mathrm{d}}(t) \, M_{\mathrm{cl}}  \, R_{\mathrm{SN}}(t),
\end{equation}
which shows that besides $R_{\mathrm{SN}}(t)$ and $M_{\mathrm{cl}}$,
the dust destruction rate in the ISM also depends on the dust-to-gas ratio $\eta_{\mathrm{d}}(t)$.

An expression for $M_{\mathrm{cl}}$ which is dependent on the shock velocity is given by \citet{dwe07}. 
For a homogenous ISM and under the assumption that silicon and carbon grains are equally mixed,
\citet{dwe07} obtains $M_{\mathrm{cl}}$ =  1100--1300 $\Msun$. 
However, the ISM is inhomogeneous, characterized by cold, warm, and hot phases with different 
densities \citep[e.g.,][and references therein]{mct89}. 
The density contrast between the cold and warm phases and the hot phase can be relatively large. 
Shocks traveling through these phases are found to be very inefficient in destroying dust \citep{jon04}. 
The destruction process is solely effective in the warm (T $>$ 100K) phase of the interstellar medium, 
while SN shocks propagating through either a hot ISM with low density or cold clouds (atomic and molecular) do not destroy dust effectively  \citep[e.g.,][]{mct89}.  

Another important parameter is the injection timescale of stellar yields and dust into the ISM.  
Following \citet{mct89}, we estimate the injection timescale of the dust from stellar sources as

\begin{equation}
\label{EQ:ITS}
           \tau_{\mathrm{in}}(t) \simeq \frac{M_{\mathrm{d}}(t)}
                                                           {E_{\mathrm{d,SN}}(t) + E_{\mathrm{d,AGB}}(t)}.
\end{equation}       
In GAH11 we made a rough estimate of the minimum averaged 
dust injection rate from SNe and AGBs of 0.5 $\Msun$ yr$^{-1}$ 
based on data for high-$z$ QSOs. 
Using this dust injection rate we obtained an average injection time $\tau_{\mathrm{in}}(t)$ = 4 $\times$ 10$^{8}$ yr.

For comparison we estimate the lifetime $\tau_{\mathrm{dl}}(t)$ of the dust grains. 
The mass of the ISM is assumed to be $M_{\mathrm{ISM}}$ = 2 $\times$ 10$^{10}$ $\Msun$. 
The SN rates are calculated for a  constant  SFR of 500 $\Msun$  yr$^{-1}$ and for a Larson 2 IMF. 
This results in a timescale of
$\tau_{\mathrm{dl}}(t)$ $\sim$ 1.3 $\times$ 10$^{7}$ yr for $M_{\mathrm{cl}}$ = 100 $\Msun$.
A shorter timescale $\tau_{\mathrm{dl}}(t)$ $\sim$ 1.6 $\times$ 10$^{6}$ yr is obtained for $M_{\mathrm{cl}}$ = 800 $\Msun$. 
When using a Salpeter IMF for the SN rates, the timescales are usually longer 
($\tau_{\mathrm{dl}}(t)$ $\sim$ 6.6 $\times$ 10$^{7}$ yr, for $M_{\mathrm{cl}}$ = 100 $\Msun$). 

We note that this is a rather rough estimate. 
Typically $\tau_{\mathrm{in}}(t)$ and $\tau_{\mathrm{dl}}(t)$ are strongly dependent on the IMF, 
$M_{\mathrm{ISM}}(t)$ and the SNe dust production efficiency $\epsilon_{\mathrm{i}}(m,Z)$. 
Hence, the values will deviate from the above estimated average during evolution. 

However, this example demonstrates that the dust injection timescale can be longer than the lifetimes of the dust grains. 
The difference between these timescales is influenced by the value of  $M_{\mathrm{cl}}$. 
Pertaining to the formation of high dust masses in galaxies, this has significant consequences.  
An injection timescale $\tau_{\mathrm{in}}(t)$ longer than the destruction timescale 
$\tau_{\mathrm{dl}}(t)$ does not allow the build up of high dust masses in the galaxy. 
This implies that the dust injection rate must be higher than the dust destruction rate. 
A lowering of the dust destruction rate could be achieved if dust grains are shielded from destruction.
Alternatively, a rapid dust grain growth in the ISM might be an option, if SNe and AGBs cannot generate the 
necessary high dust injection rate.
Grain growth, however, is not incorporated into our model and remains to be investigated.

Furthermore, we assume that the starburst occurs in an initially dust free galaxy. 
Consequently dust produced by the first generations of SNe might be 
ejected into the ISM unhindered. 
The epoch at which the first SN shocks are able to sweep up ISM gas mixed with dust remains elusive. 
In view of these considerations, we stress that dust destruction in the ISM is uncertain, 
particularly when considering galaxies with conditions as described in Sect.~\ref{SSC:ICG} and \ref{SSC:IMF}.

Despite our simple assumption of dust being immediately homogeneously mixed with the gas in the ISM, 
we account for the uncertainty of the lifetime of dust grains against destruction 
by using $M_{\mathrm{cl}}$  as a parameter. 
This has also been done by \citet{dwe07}.
The considered cases are for $M_{\mathrm{cl}}$ = 800 $\Msun$ as the highest destruction, 
$M_{\mathrm{cl}}$ = 100 $\Msun$ for modest destruction, and $M_{\mathrm{cl}}$ = 0  for no dust destruction. 
%
%
%
\section{Results}
\label{SEC:RES}
%
In this section we present the results of several models calculated within the first Gyr after starburst. 
We focus on the total dust mass in a galaxy.
At redshift $z$ $>$ 6, the time to build up high dust masses of $>$ 10$^{8}$ $\Msun$ is limited to 400--500 Myr. 
Models able to exceed an amount of 10$^{8}$ $\Msun$ of dust within this time are therefore of particular interest.  
General evolutionary tendencies of certain quantities, such as dust injection rates, SFR, metallicity, 
or the amount of gas, are also discussed.
In this paper we present results that assume the same value for the initial SFR in all models as given in Table~\ref{TAB:IPP}.
In \citet{gall10c}, we investigated models for different initial SFRs and discuss their applicability to particular quasars at $z$ $\gtrsim$ 6. 
%
\subsection{Evolution of dust, dust-to-gas, and dust-to-metal ratios}
%
We study models with all considered initial gas masses $M_{\mathrm{ini}}$.  
The models discussed are calculated for the case that no SN occurs when a BH is formed (see Sect.~\ref{SSS:RGM}).
Models including SMBH formation are deferred to Sect.~\ref{SEC:MSMBH}.

\subsubsection{Models with SN type differentiation}
%
    \begin{figure*}
   \centering
   \includegraphics[width=\textwidth]{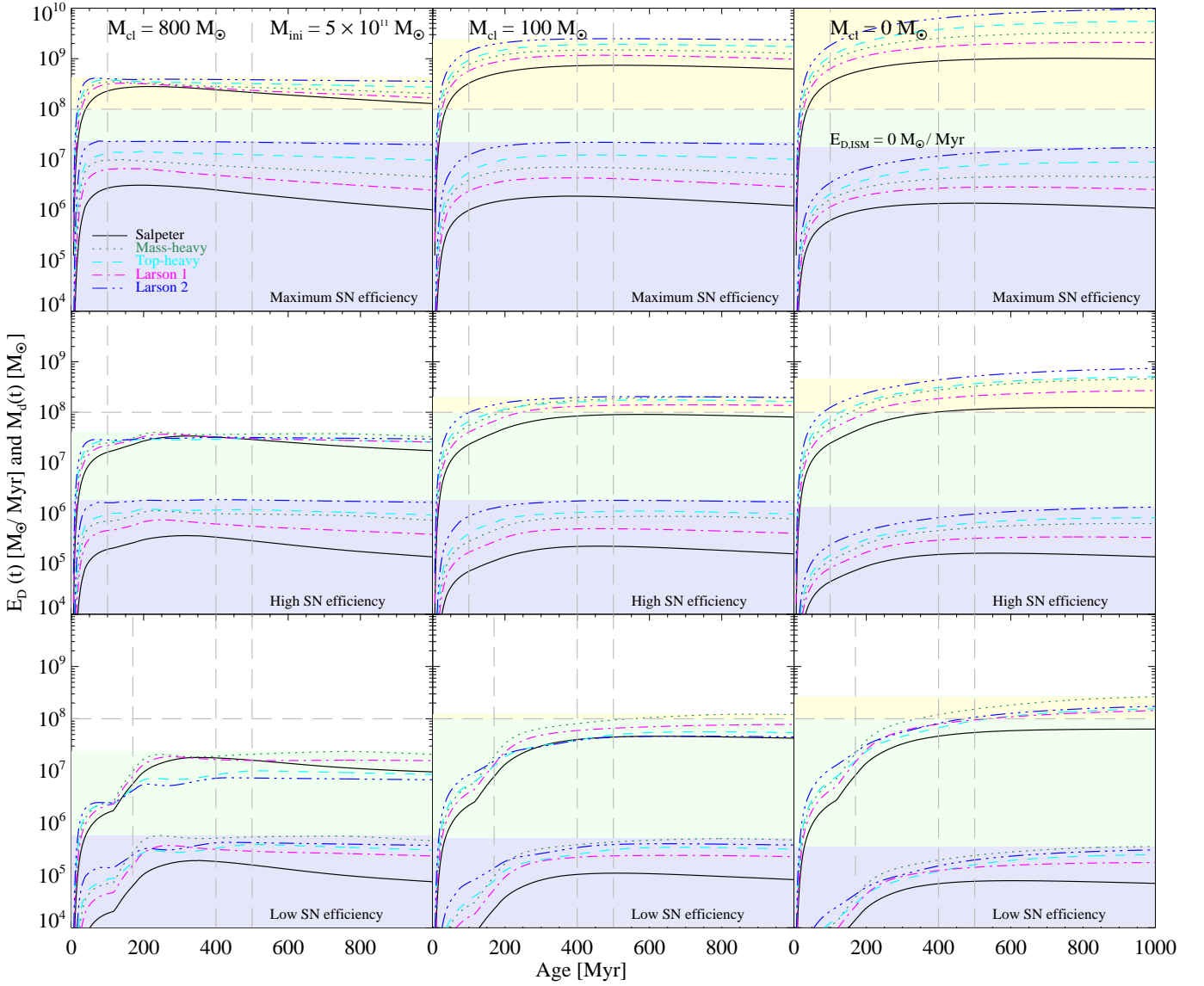}   
      \caption{Evolution of the total dust mass and dust destruction rates for EIT08M. 
      The initial gas mass of the galaxy $M_{\mathrm{ini}}$ = 5 $\times$ 10$^{11}$ $\Msun$. 
      Calculations are performed for a `maximum' (top row),
      a `high' (middle row), and a `low' (bottom row) SN dust production efficiency $\epsilon_{\mathrm{i}}(m)$. 
      Dust destruction is taken into account for 
      $M_{\mathrm{cl}}$ = 800 $\Msun$ (left column), 100 $\Msun$ (middle column), and 0 $\Msun$ (right column). 
      The bottom group of curves in the light blue area represents the total dust destruction rate $E_{\mathrm{D}}(t)$, 
       which is based on Eq.~\ref{EQ:DDEST}.
       The upper group of curves in the green or yellow zones displays 
      the evolution of the total amount of dust $M_{\mathrm{d}}(t)$ in the galaxy. 
      The yellow area marks dust masses exceeding $10^8$ $\Msun$ 
      of dust. The grey horizontal dashed line marks the limit of $10^8$ $\Msun$ of dust. 
      The grey vertical dashed lines indicate epochs at 100, 
      170, 400, and 500 Myr after the onset of starburst. 
      The black solid, green dotted, cyan dashed, magenta dashed-dotted, and 
      blue dashed-dot-dotted lines represent the Salpeter, mass-heavy, top-heavy, 
      Larson 1 and Larson 2 IMF, respectively.  
                      }
     \label{FIG:DUSE}
   \end{figure*}
    \begin{figure*}
   \centering
   \includegraphics[width=\textwidth]{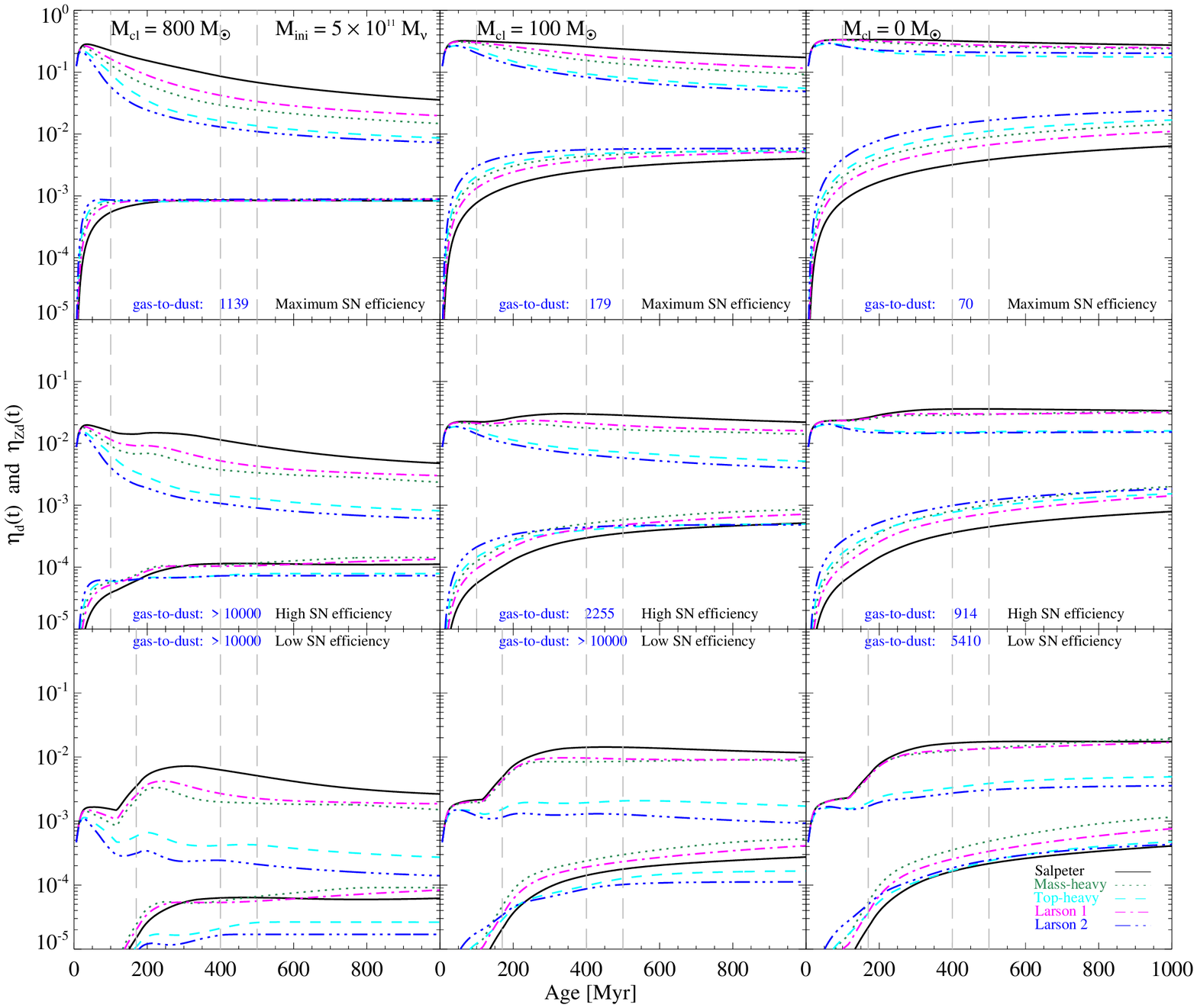}
      \caption{Evolution of the dust-to-gas and dust-to-metal mass ratios for EIT08M. 
      The initial gas mass of the galaxy 
      $M_{\mathrm{ini}}$ = 5 $\times$ 10$^{11}$ $\Msun$. 
      Calculations are performed for a `maximum' (top row),
      a `high' (middle row), and a `low' (bottom row) SN dust production efficiency $\epsilon_{\mathrm{i}}(m)$. 
      Dust destruction is taken into account for $M_{\mathrm{cl}}$ = 800 $\Msun$ (left column), 
      100 $\Msun$ (middle column), and 0 $\Msun$ (right column).
      The upper group of curves signifies the dust-to-metal mass ratio $\eta_{\mathrm{Zd}}(t)$. 
      The lower group of curves represents the dust-to-gas mass ratio $\eta_{\mathrm{d}}(t)$. 
      The gas-to-dust ratio displayed in the figures is calculated for a Larson 2 IMF at an epoch
      of 400 Myr. 
      The grey vertical dashed lines indicate epochs at 100, 400, and 500 Myr after the onset of starburst.
      The black solid, green dotted, cyan dashed, magenta dashed-dotted, 
      and blue dashed-dot-dotted lines represent the Salpeter, 
      mass-heavy, top-heavy, Larson 1 and Larson 2 IMF, respectively. 
                          }
     \label{FIG:DUSE_R}                    
   \end{figure*}
    \begin{figure*}
   \centering
   \includegraphics[width=\textwidth]{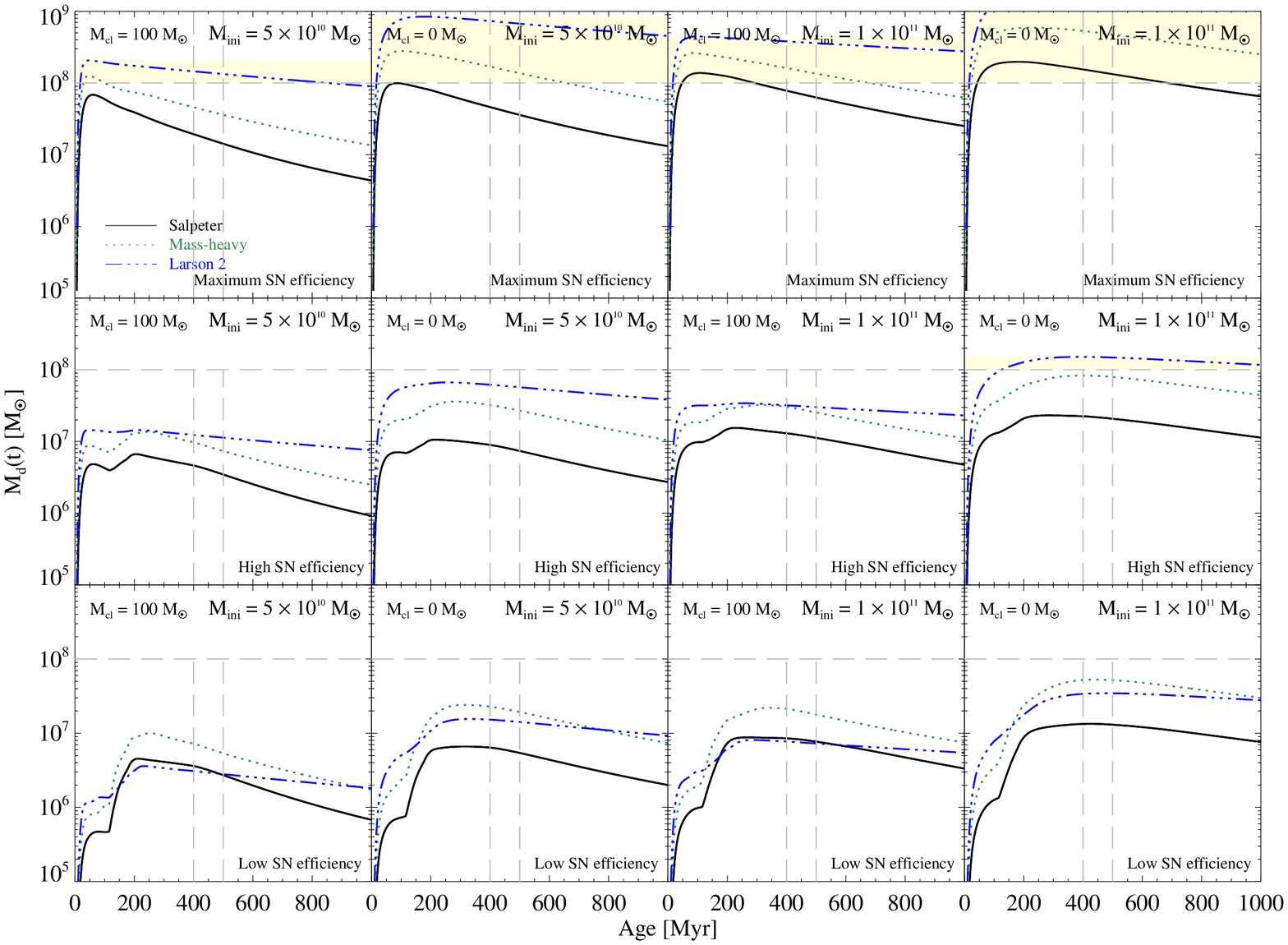}   
      \caption{Evolution of the total dust mass for EIT08M. 
      Calculations for galaxies with initial gas masses of 
      $M_{\mathrm{ini}}$ = 5 $\times$ 10$^{10}$ $\Msun$ are presented in the 
      first and second columns and for 
      $M_{\mathrm{ini}}$ = 1 $\times$ 10$^{11}$ $\Msun$ in the third and fourth columns. 
      Total dust masses $M_{\mathrm{d}}(t)$ are shown for a `maximum' (top row), a `high' (middle row), 
      and a `low' (bottom row) SN dust production efficiency $\epsilon_{\mathrm{I}}(m)$. 
      Dust destruction is taken into account for  $M_{\mathrm{cl}}$ =  100 $\Msun$ 
      (first and third columns) and 0 $\Msun$ (second and fourth columns). 
      Maximum dust masses  exceeding $10^8$ $\Msun$ are indicated 
      as yellow shaded zones.  
       The grey horizontal dashed line marks the limit of $10^8$ $\Msun$ of dust. 
       The grey vertical dashed lines indicate epochs at 
      400, and 500 Myr after the onset of starburst. 
      The black solid, green dotted, and 
      blue dashed-dot-dotted lines represent the Salpeter, mass-heavy 
      and Larson 2 IMF, respectively.  
                                          }
     \label{FIG:DUSE15}
   \end{figure*}
    \begin{figure*}
   \centering
   \includegraphics[width=\textwidth]{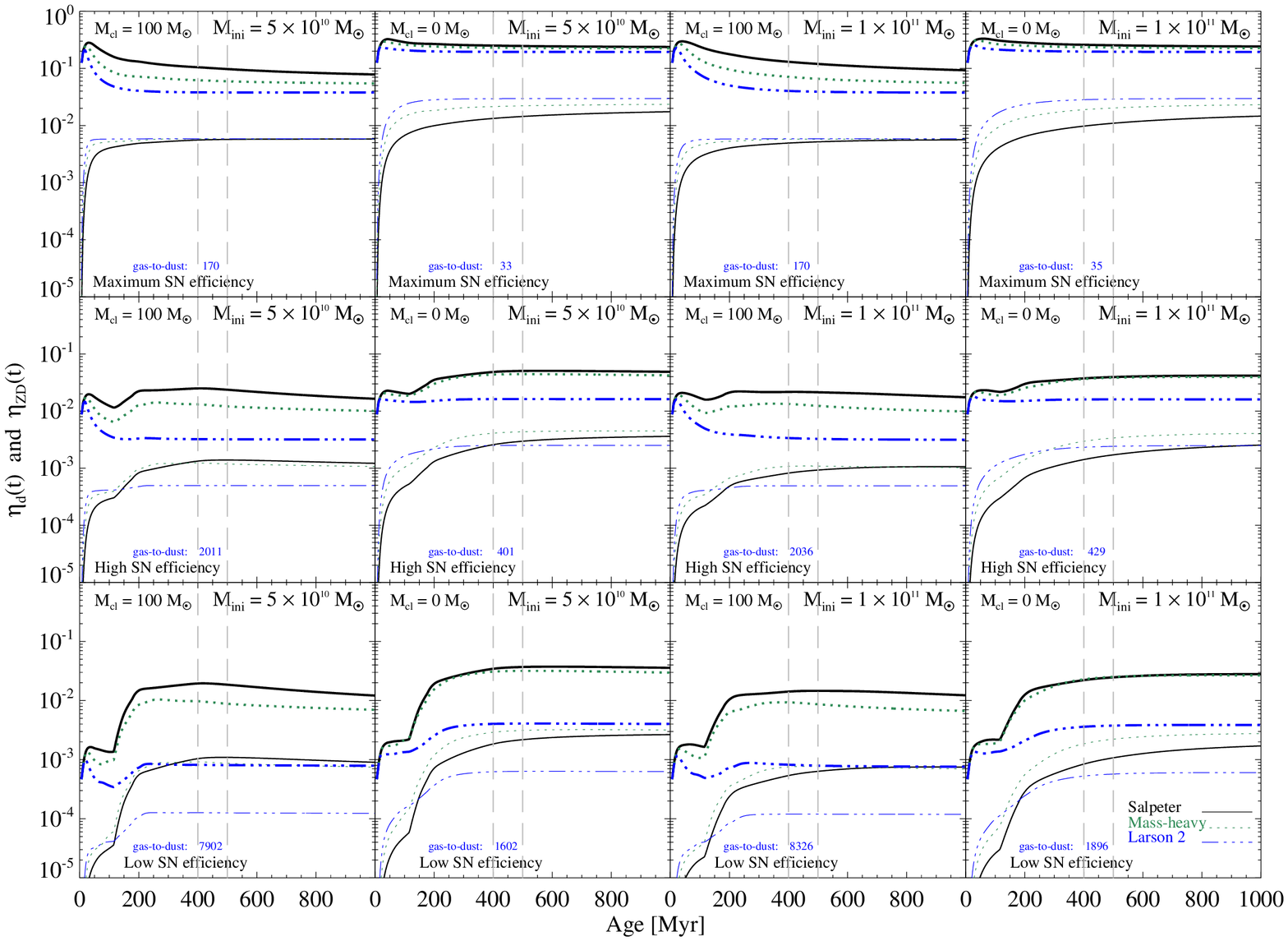}   
      \caption{Evolution of the dust-to-gas and dust-to-metal mass ratios for EIT08M.
      Calculations for galaxies with initial gas masses of 
      $M_{\mathrm{ini}}$ = 5 $\times$ 10$^{10}$ $\Msun$ are presented in the 
      first and second columns  
      and for $M_{\mathrm{ini}}$ = 1 $\times$ 10$^{11}$ $\Msun$ in the third and fourth columns. 
      Results are presented for a `maximum' (top row), a `high' (middle row), and a `low' 
      (bottom row) SN dust production efficiency $\epsilon_{\mathrm{i}}(m)$. 
      Dust destruction is taken into account for $M_{\mathrm{cl}}$ =  100 $\Msun$ 
      (first and third columns) and 0 $\Msun$ (second and fourth columns). 
      The upper group of  thick curves signifies the dust-to-metal mass ratio $\eta_{\mathrm{Zd}}(t)$. 
      The lower group of  thin curves represents the dust-to-gas mass ratio $\eta_{\mathrm{d}}(t)$. 
      The gas-to-dust ratio displayed in the figures is calculated for a Larson 2 IMF at an epoch
      of 400 Myr. 
      The grey vertical dashed lines indicate epochs at 400, and 500 Myr after the onset of starburst.
      The black solid, green dotted,
      and blue dashed-dot-dotted lines represent the Salpeter, 
      mass-heavy and Larson 2 IMF, respectively. 
                                    }
     \label{FIG:DUSE15_R}
   \end{figure*}
\paragraph{EIT08M with $M_{\mathrm{ini}}$ = 5 $\times$  $10^{11}$ $\Msun$:}
In Fig.~\ref{FIG:DUSE} the results are presented for the mass of dust $M_{\mathrm{d}}(t)$ 
in a galaxy with $M_{\mathrm{ini}}$ = 5 $\times$  $10^{11}$ $\Msun$. 
Calculations were performed for a `maximum' (top row), a  `high' (middle row),  and a `low' (bottom row) 
SN dust production efficiency $\epsilon_{\mathrm{i}}(m)$.  
Dust destruction in the ISM is included for three values of $M_{\mathrm{cl}}$ (see Sect. \ref{SSC:DISM}). 
This model is considered as a reference model. We therefore describe it in detail.   

Dust masses $M_{\mathrm{d}}(t)$ obtained with the `maximum' SN dust production efficiency 
$\epsilon_{\mathrm{max}}(m)$ exceed $10^8$ $\Msun$ of dust
in all cases of $M_{\mathrm{cl}}$ and for all IMFs. 
In fact, $10^8$ $\Msun$ of $M_{\mathrm{d}}(t)$ is reached within the first few Myr. 
This is insensitive to both the IMF and the amount of  $M_{\mathrm{cl}}$. 
The latter however determines the suppression of $M_{\mathrm{d}}(t)$ as the system evolves. 
For  $M_{\mathrm{cl}}$ = 800 $\Msun$ (top left), a maximum amount of $\sim$ 4 $\times$ $10^8$ $\Msun$ 
is reached for a Larson 2 IMF and sustained until the end of the computation. 
Only the IMFs favouring lower masses exhibit a shallow decline in  $M_{\mathrm{d}}(t)$.  
Dust destruction with $M_{\mathrm{cl}}$ = 100 $\Msun$ (top middle) leads to nearly constant 
dust masses retainable over 1 Gyr of evolution. 
The amount of dust for most IMFs is a few times $10^9$ $\Msun$. 
When assuming no dust destruction (top right), a broader spread of the dust masses 
for the different IMFs develops. 
The amount of dust increases with time regardless of the IMF and 
yields up to $10^{9-10}$ $\Msun$ are reached. 
  
In the case of `high' SN dust production efficiency, $\epsilon_{\mathrm{high}}(m)$, similar trends are featured.  
However the amount of dust is lower.  
Without dust destruction ($M_{\mathrm{cl}}$ = 0), dust masses up to a few times 
$10^8$ $\Msun$ are attained for all IMFs (Fig.~\ref{FIG:DUSE} middle right). 
The timescale to exceed $10^8$ $\Msun$ of dust ranges from 80 Myr (Larson 2 IMF) up to 400 Myr (Salpeter IMF). 
Taking dust destruction with only a modest amount of $M_{\mathrm{cl}}$ = 100 $\Msun$ 
into account decreases $M_{\mathrm{d}}(t)$ substantially (Fig.~\ref{FIG:DUSE} middle middle). 
The IMFs favouring high masses are most affected, while the reduction in $M_{\mathrm{d}}(t)$ 
for a Salpeter IMF is small. 
Except for the Salpeter IMF, all remaining IMFs lead to more than $10^8$ $\Msun$ of dust. 
The highest amount of dust is reached with either a Larson 2 or a mass-heavy IMF, and is 
$\sim$ 2 $\times$  $10^8$ $\Msun$ at an epoch of $\sim$ 400 Myr. 
Considering a very high destruction with $M_{\mathrm{cl}}$ = 800 $\Msun$ results in a strong 
reduction of  $M_{\mathrm{d}}(t)$, so only $\sim$ 2--3 $\times$  $10^7$ $\Msun$ of dust 
remains throughout the evolution (Fig.~\ref{FIG:DUSE} middle left). 

The results for $M_{\mathrm{d}}(t)$ with a `low' SN dust production efficiency 
$\epsilon_{\mathrm{low}}(m)$ are dominated by AGB dust production (see also  GAH11). 
Without dust destruction, dust masses up to 1--2 $\times$ 10$^8$ $\Msun$ is reached 
after $\sim$ 300--500 Myr for most of the IMFs. 
Applying a modest dust destruction of $M_{\mathrm{cl}}$ = 100 $\Msun$ reduces 
$M_{\mathrm{d}}(t)$ analogously to the higher SN efficiencies  
$\epsilon_{\mathrm{max}}(m)$ or $\epsilon_{\mathrm{high}}(m)$. 
Dust masses for either the top-heavy or the Larson 2 IMF are efficiently decreased and 
comparable to $M_{\mathrm{d}}(t)$ for a Salpeter IMF.
Increasing $M_{\mathrm{cl}}$ to 800 $\Msun$ leads to a stronger reduction in 
$M_{\mathrm{d}}(t)$ for these IMFs, resulting in lower dust masses than for a Salpeter IMF. 

A considerable difference in the progression of $M_{\mathrm{d}}(t)$ for the `low' SN efficiency 
$\epsilon_{\mathrm{low}}(m)$ compared to the higher SN efficiencies 
is encountered during the first $\sim$ 200 Myr. 
While a fast rise of $M_{\mathrm{d}}(t)$ is identifiable
for either $\epsilon_{\mathrm{max}}(m)$ or $\epsilon_{\mathrm{high}}(m)$, 
for $\epsilon_{\mathrm{low}}(m)$
the dust mass remains between $10^{6-7}$ $\Msun$ during this epoch. 
This is caused primarily by high-mass AGB stars with short lifetimes ($>$ 4--5 $\Msun$). 
A further increase in the dust mass due to the delayed AGB dust injection from the less massive 
but more efficient AGB stars results in a second dust bump at an epoch of $\sim$ 200--300 Myr. 
High dust masses at early epochs (100--200 Myr) are not possible with $\epsilon_{\mathrm{low}}(m)$.

The curves at the bottom of the light-blue area in Fig.~\ref{FIG:DUSE} represent the 
dust destruction rate $E_{\mathrm{D}}(t)$.
For $M_{\mathrm{cl}}$ = 0  and independent of $\epsilon_{\mathrm{i}}(m)$ the dust destruction rate 
reflects the amount of dust incorporated into stars per unit time and is calculated as 
$E_{\mathrm{D}}(t) = \eta_{\mathrm{d}}(t) \, \psi(t)$.
For  $M_{\mathrm{cl}} > 0$ $E_{\mathrm{D}}(t)$ is given by Eq.~\ref{EQ:DDEST} and 
additionally consists of the destroyed dust in the ISM, $E_{\mathrm{D,ISM}}(t)$.

We find that $E_{\mathrm{D}}(t)$ increases faster and earlier with increasing $M_{\mathrm{cl}}$. 
Consequently, dust destruction rates comparable to SN injection rates are reached at earlier epochs.   
Given  Eq.~\ref{EQ:DESTR} for the amount of dust destroyed through SN shocks, 
 the high SN rates $R_{\mathrm{SN}}(t)$ at the beginning of evolution in conjunction with 
 $M_{\mathrm{cl}}$ lead to a high base value of $E_{\mathrm{D,ISM}}(t)$.  
This suppresses the rise in the amount of dust at early epochs. 
Additionally, SN rates are higher for the IMFs biased towards more massive stars affecting these IMFs most. 
Later in the evolution, the dust destruction rates remain almost constant in most cases, with a marginal increase 
for $M_{\mathrm{cl}}$ = 0 or decline for $M_{\mathrm{cl}}$ $>$ 0.  
Dust destruction and injection appear to be balanced, resulting in the flat development of $M_{\mathrm{d}}(t)$.

In Fig. \ref{FIG:DUSE_R} we present the results for the evolution of the dust-to-gas mass ratio 
$\eta_{\mathrm{d}}(t)$ and the dust-to-metal mass ratio $\eta_{\mathrm{Zd}}(t)$. 
Commonly the curves for $\eta_{\mathrm{d}}(t)$ for all IMFs are slowly increasing with time 
except for $M_{\mathrm{cl}}$ = 800 $\Msun$ for which they remain approximately constant. 
Interestingly, in this case and for $\epsilon_{\mathrm{max}}(m)$, the dust-to-gas ratio for all IMFs 
sustains a constant value of $\sim$ $10^{-3}$ over the whole evolution. 
Without dust destruction (top right), $\eta_{\mathrm{d}}(t)$ increases up to values above 
$\sim$ $10^{-2}$ for most of the IMFs. 
For either a `high' or `low' SN efficiency, the values are below $\sim$ $10^{-3}$. 
Only for $\epsilon_{\mathrm{high}}(m)$ and $M_{\mathrm{cl}}$ = 0  (middle right) 
is $\eta_{\mathrm{d}}(t)$ further increased.
These overall  trends show that  $\eta_{\mathrm{d}}(t)$ 
can be very low for massive galaxies even if the galaxy appears dusty.

The dust-to-metal mass ratio $\eta_{\mathrm{Zd}}(t)$ is significantly lower for the Larson 2 and
top-heavy IMF than for the other IMFs. 
This feature is exhibited in all calculated models. 
The degree of this separation between different IMFs depends primarily on the SN 
dust production efficiency and 
secondly on the destruction rate in the ISM, and is larger for $\epsilon_{\mathrm{low}}(m)$.   
This reflects  the decrease in the SN dust production efficiency towards the higher 
mass end of the SN mass interval. 
Additionally, Type Ib/c SNe produce no dust, but they do inject metals into the ISM 
at higher rates for the top-heavy IMFs. 
For $M_{\mathrm{cl}}$ = 0  and $\epsilon_{\mathrm{high}}(m)$, the amount of metals 
bound in dust grains is between $\sim$ 1.5--3 \%. 
For $M_{\mathrm{cl}}$ = 100 $\Msun$  the fraction of metals bound in dust grains for the two IMFs 
favouring high masses is below 1 \% and decreases further with increasing $M_{\mathrm{cl}}$. 
Generally, for a `low' SN  efficiency, $\eta_{\mathrm{Zd}}(t)$ remains below $10^{-2}$ for all IMFs.
For the `maximum' SN efficiency $\epsilon_{\mathrm{max}}(m)$ the dust-to-metal ratio, $\eta_{\mathrm{Zd}}(t)$, 
is about a factor of 10 higher than for $\epsilon_{\mathrm{high}}(m)$.   

\paragraph{EIT08M with $M_{\mathrm{ini}}$ = 1 $\times$  $10^{11}$ $\Msun$:}
In Fig.~\ref{FIG:DUSE15} (two right columns) the results are presented for 
$M_{\mathrm{d}}(t)$ in cases of either no or a modest destruction and all three 
SN dust formation efficiencies $\epsilon_{\mathrm{max}}(m)$, 
$\epsilon_{\mathrm{high}}(m)$, $\epsilon_{\mathrm{low}}(m)$. 

The top row of Fig.~\ref{FIG:DUSE15} depicts the evolution of  the amount of dust 
$M_{\mathrm{d}}(t)$ for $\epsilon_{\mathrm{max}}(m)$. 
It is evident that high dust masses beyond $10^8$  $\Msun$ are obtained for both cases of $M_{\mathrm{cl}}$.  
The maximum value for $M_{\mathrm{d}}(t)$ is already reached shortly after the onset of the starburst.  
Thereafter $M_{\mathrm{d}}(t)$ follows a negative slope, which is steepest for the IMFs biased towards low-mass stars. 
This decline is generally observable in all cases of $M_{\mathrm{cl}}$ and $\epsilon_{\mathrm{i}}(m)$.
The cause of this is that dust is incorporated into stars on higher rates than is replenished by stellar sources. 
This dust decrease is amplified by dust destruction in the ISM as is demonstrated for $M_{\mathrm{cl}} = 100$ $\Msun$.  
For $\epsilon_{\mathrm{high}}(m)$ and $M_{\mathrm{cl}}$ = 0,  a high total dust mass $\ge$ $10^{8}$ $\Msun$ 
can be achieved with a Larson 2 IMF after $\sim$ 200 Myrs (Fig.~\ref{FIG:DUSE15} middle row, right column). 
For all remaining cases of either $M_{\mathrm{cl}}$ or for $\epsilon_{\mathrm{low}}(m)$ and 
regardless of the IMF, $M_{\mathrm{d}}(t)$ stays below $10^{8}$ $\Msun$. 
The tendencies for the various IMFs resemble those identified for $M_{\mathrm{ini}}$ = 5 $\times$  $10^{11}$ $\Msun$. 

The curves of the dust-to-gas ratio $\eta_{\mathrm{d}}(t)$ and the dust-to-metal ratio 
$\eta_{\mathrm{ZD}}(t)$ are also similar to the system with 
$M_{\mathrm{ini}}$ = 5 $\times$  $10^{11}$ $\Msun$, 
although the amount is higher (see Fig. \ref{FIG:DUSE15_R} two left columns). 
For $M_{\mathrm{cl}}$ = 0  and `maximum' SN efficiency, $\eta_{\mathrm{d}}(t)$ is roughly 
independent of the IMF, while exhibiting a wider range for $\epsilon_{\mathrm{low}}(m)$.
The fraction of metals bound in dust grains is 
 $\lesssim$ 4 \% ($M_{\mathrm{cl}}$ = 0) for IMFs favouring low-mass stars
and $\epsilon_{\mathrm{high}}(m)$; 
 for $\epsilon_{\mathrm{max}}(m)$ it is roughly 20--30 \%.

\paragraph{EIT08M with $M_{\mathrm{ini}}$ = 5 $\times$  $10^{10}$ $\Msun$:}
The evolution of $M_{\mathrm{d}}(t)$ within the first Gyr exhibits similar, 
but more strongly pronounced, tendencies for the various IMFs as 
the galaxy system with $M_{\mathrm{ini}}$ = 1 $\times$  $10^{11}$ $\Msun$. 
Results are presented in Fig.~\ref{FIG:DUSE15} (two left columns). 
Even with a `maximum' SN efficiency $\epsilon_{\mathrm{max}}(m)$, 
large dust masses of a few times $10^{8}$ $\Msun$ cannot be reached or sustained 
when dust destruction in the ISM is considered.  
In Fig.~\ref{FIG:DUSE15_R} (two left columns) it is evident that the dust-to-gas ratios 
$\eta_{\mathrm{d}}(t)$ and the dust-to-metal ratios $\eta_{\mathrm{ZD}}(t)$ result in 
higher values than for the higher mass galaxy systems 
(see Figs. \ref{FIG:DUSE_R}, \ref{FIG:DUSE15_R} two right columns). 
The tendencies are the same in general.  
\paragraph{G09M:}
    \begin{figure}
   \centering
   \resizebox{\hsize}{!}{ \includegraphics{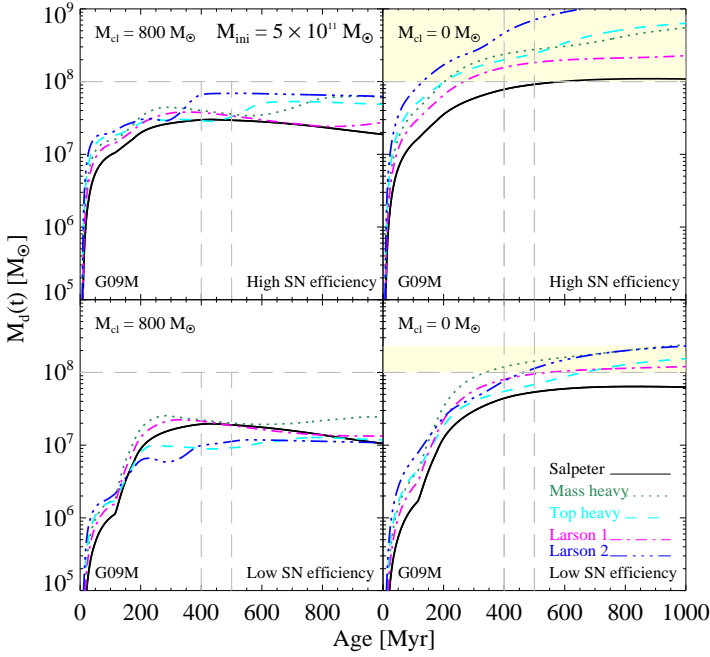}}   
      \caption{Evolution of the total dust mass for G09M,
      presented for a galaxy with 
      $M_{\mathrm{ini}}$ = 5 $\times$ 10$^{11}$ $\Msun$. 
      Total dust masses $M_{\mathrm{d}}(t)$ are shown for a `high' (top row) 
      and a `low' (bottom row) SN dust production efficiency $\epsilon_{\mathrm{i}}(m)$. 
      Dust destruction is taken into account for $M_{\mathrm{cl}}$ =  800 $\Msun$ 
      (left column) and 0 $\Msun$ (right column). 
      Maximum dust masses  exceeding $10^8$ $\Msun$ are indicated 
      as yellow shaded zones.  
       The grey horizontal dashed line marks the limit of $10^8$ $\Msun$ of dust. 
       The grey vertical dashed lines indicate epochs at 
      400 and 500 Myr after the onset of starburst. 
      The black solid, green dotted, cyan dashed, magenta dashed-dotted, and 
      blue dashed-dot-dotted lines represent the Salpeter, mass-heavy, top-heavy, 
      Larson 1, and Larson 2 IMF, respectively.  
                                          }
     \label{FIG:G09M}
   \end{figure}
In Fig.~\ref{FIG:G09M} 
the results are shown for models with a rotationally enhanced mass-loss prescription. 
The galaxy under consideration has $M_{\mathrm{ini}}$ = 5 $\times$  $10^{11}$ $\Msun$. 
The outcome is similar to the reference EIT08M (Fig.~\ref{FIG:DUSE}), 
although the obtained dust yields are increased for the IMFs favouring higher masses. 
For a Larson 2 IMF, $M_{\mathrm{d}}(t)$ exceeds $10^{9}$ $\Msun$ after $\sim$ 650 Myr. 
Interestingly, even though the dust yields are high for no dust destruction in the ISM, 
for $M_{\mathrm{cl}}$ = 800 $\Msun$ the dust mass also stays below $10^{8}$ $\Msun$. 
In case of a `low' SN dust production efficiency, $10^{8}$ $\Msun$ 
cannot be reached within the first 400 Myr. 
However, for the IMFs biased towards high masses an amount of dust $>$ $10^{8}$ $\Msun$ is attained later in the evolution. 
\subsubsection{Models with fixed SNe mass range}  
%
    \begin{figure}
   \centering
   \resizebox{\hsize}{!}{ \includegraphics{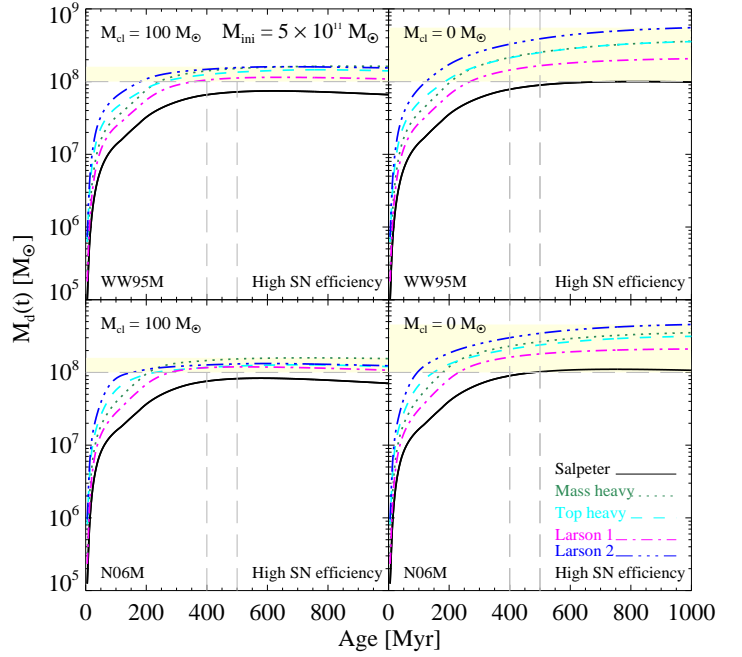}}   
      \caption{ Evolution of the total dust mass for WW95M (top row) and N06M (bottom row).
      Both models are presented for a galaxy with $M_{\mathrm{ini}}$ = 5 $\times$ 10$^{11}$ $\Msun$. 
      Total dust masses $M_{\mathrm{d}}(t)$ are shown for `high' SN dust production 
      efficiency $\epsilon_{\mathrm{high}}(m)$. 
      Dust destruction is taken into account for $M_{\mathrm{cl}}$ =  100 $\Msun$ (left column) 
      and 0 $\Msun$ (right column). 
      Maximum dust masses  exceeding $10^8$ $\Msun$ are indicated 
      as yellow shaded zones.  
       The grey horizontal dashed line marks the limit of $10^8$ $\Msun$ of dust. 
       The grey vertical dashed lines indicate epochs at 
      400 and 500 Myr after the onset of starburst. 
      The black solid, green dotted, cyan dashed, magenta dashed-dotted, and 
      blue dashed-dot-dotted lines represent the Salpeter, mass-heavy, top-heavy, 
      Larson 1, and Larson 2 IMF, respectively.  
                                          }
     \label{FIG:PLEVOLNOMO}
   \end{figure}
In Fig.~\ref{FIG:PLEVOLNOMO} the results of $M_{\mathrm{d}}(t)$ are shown for 
WW95M (top row) and N06M (bottom row) for a galaxy with 
$M_{\mathrm{ini}}$ = 5 $\times$  $10^{11}$ $\Msun$ and `high' SN efficiency $\epsilon_{\mathrm{high}}(m)$.  
The amount of dust in N06M early in the evolution 
is greater than for WW95M for the IMFs favouring massive stars, but
flattens later, leading 
to slightly lower dust masses than achieved with WW95M.  
The dust masses achieved with either the `maximum' or the `low' SN dust production efficiencies 
are nearly identical to EIT08M and are therefore not shown. 
Generally, the evolution of these models is similar to EIT08M.
%
%
\subsubsection{ Very high $M_{\mathrm{ini}}$ and the case of constant SFR}
\label{SSC:CSFR}
%
    \begin{figure}
   \centering
   \resizebox{\hsize}{!}{ \includegraphics{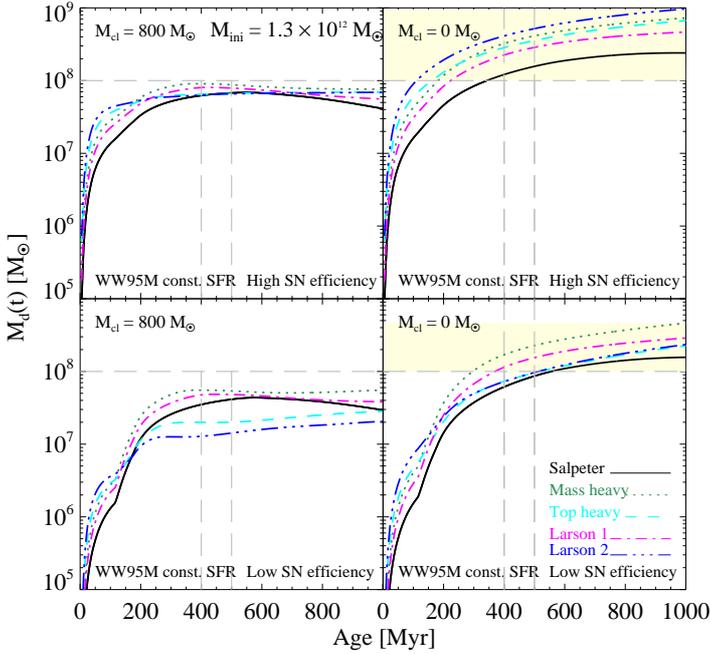}}   
      \caption{Evolution of the total dust mass for WW95M with constant SFR.
      WW95M is presented for a galaxy with 
      $M_{\mathrm{ini}}$ = 1.3 $\times$ 10$^{12}$ $\Msun$. 
      Total dust masses $M_{\mathrm{d}}(t)$ are shown for `high' (top row) and 
      a `low' (bottom row) SN dust production efficiency $\epsilon_{\mathrm{i}}(m)$. 
      Dust destruction is taken into account for $M_{\mathrm{cl}}$ =  800 $\Msun$ 
      (left column) and 0 $\Msun$ (right column). 
      Maximum dust masses  exceeding $10^8$ $\Msun$ are indicated 
      as yellow shaded zones.  
       The grey horizontal dashed line marks the limit of $10^8$ $\Msun$ of dust. 
       The grey vertical dashed lines indicate epochs at 
      400 and 500 Myr after the onset of starburst. 
      The black solid, green dotted, cyan dashed, magenta dashed-dotted, and 
      blue dashed-dot-dotted lines represent the 
      Salpeter, mass-heavy, top-heavy, Larson 1, and Larson 2 IMF, respectively.  
                                          }
     \label{FIG:WW95CONST}
   \end{figure}
In Fig.~\ref{FIG:WW95CONST} the results are displayed for a case with constant SFR 
$\psi(t)$ = 1 $\times$ $10^{3}$ $\Msun$  yr$^{-1}$ and 
a galaxy with a high initial gas mass of $M_{\mathrm{ini}}$ = 1.3 $\times$ 10$^{12}$ $\Msun$. 
Such a high $M_{\mathrm{ini}}$ is necessary when assuming a constant SFR, 
since in the lower mass galaxies the available mass for star formation gets exhausted 
before an age of 1 Gyr is reached.

We find that a build up of high dust masses 
also for such massive galaxies is greatly suppressed with a high amount of destruction. 
With none of the IMFs, $10^8$ $\Msun$ of dust is attained, apart from the models with $\epsilon_{\mathrm{max}}$. 
High dust masses with either $\epsilon_{\mathrm{high}}$ or $\epsilon_{\mathrm{low}}$ 
are only possible in case of modest or no dust destruction. 
For the latter the results are shown in the right column of  Fig.~\ref{FIG:WW95CONST}.  
Interestingly, also for `low' SN dust production efficiency, 
dust masses $>$ $10^8$ $\Msun$ are possible after $\sim$ 300--400 Myr.   

The results are very similar to an EIT08M with such high $M_{\mathrm{ini}}$ and an evolving SFR. 
A slight difference appears for $M_{\mathrm{d}}(t)$ and IMFs that favour low-mass stars.
These models remain close to constant after 500--600 Myr, 
while the models for constant SFR slowly decline.  
The similarity between these models can be explained by the very high mass of the galaxy.  
In neither model does the ISM mass get significantly reduced. 
%
%
\subsection{Evolution of dust production rates, SFR and metallicity}
\label{SSS:EDRSM} 
\begin{figure*}
   \centering
   \includegraphics[width=\textwidth]{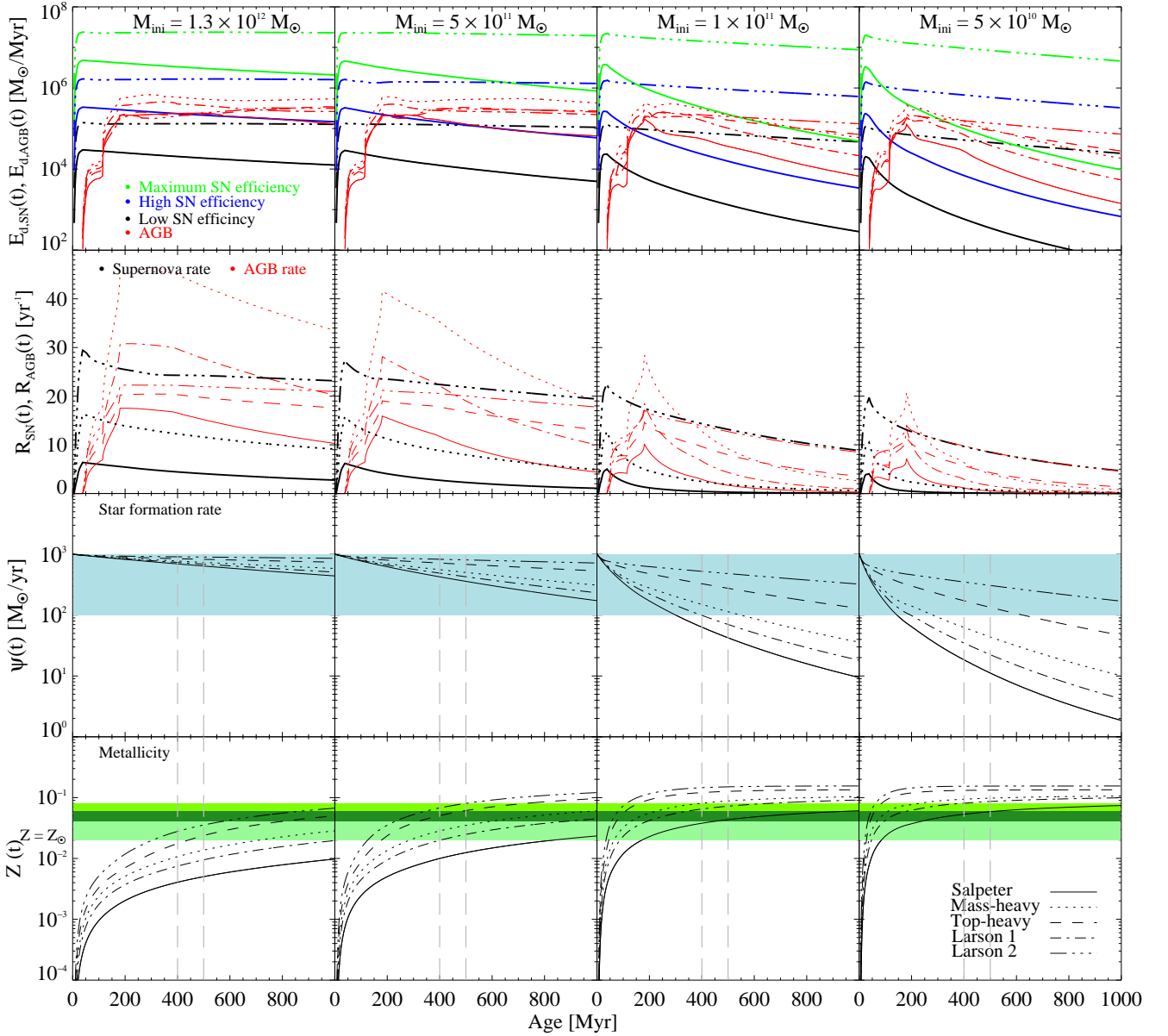}
      \caption{Evolution of total dust injection rates for AGB stars and SNe, 
      the AGB and SNe rates, the SFR, and the metallicity.
      Results are shown for $M_{\mathrm{ini}}$ = 1.3 $\times$ 10$^{12}$ $\Msun$ (first column), 
      5 $\times$ 10$^{11}$ $\Msun$ (second column),
      1 $\times$ 10$^{11}$ $\Msun$ (third column), and 
      5 $\times$ 10$^{10}$ $\Msun$ (fourth column). 
      The solid, dotted, dashed, dashed-dotted, and dashed-dot-dotted lines represent 
      the Salpeter, mass-heavy, top-heavy, Larson 1, and Larson 2 IMF, respectively. 
      First row:  Total SN dust injection rates $E_{\mathrm{d,SN}}(t)$ for a `low' (black lines), `high' (blue lines), 
      and  `maximum' (green lines) SN efficiency $\epsilon_{\mathrm{i}}(m)$, 
      and AGB dust injection rates $E_{\mathrm{d,AGB}}(t)$ (red lines). 
       SN dust injection rates are only shown for a Salpeter (solid) and a Larson 2 (dashed-dot-dotted) IMF. 
      Second row: SNe rates  $R_{\mathrm{SN}}(t)$ (black lines) and AGB rates $R_{\mathrm{AGB}}(t)$ (red lines).
      SN rates are only shown for a Salpeter (solid), mass-heavy (dotted), and a Larson 2 (dashed-dot-dotted) IMF. 
      Third row: Evolution of the SFR. The blue area marks the region between a SFR of 100--1000 $\Msun$ yr$^{-1}$. 
      Fourth row: Evolution of the metallicity $Z(t)$. The green regions mark a metallicity between 1--2 $\Zsun$ (light green), 
      2--3 $\Zsun$ (dark green), 3--4 $\Zsun$ (grass green).
      The grey vertical dashed lines indicate epochs at 400 and 500 Myr after the onset of starburst. 
              }
     \label{FIG:DUSE123}
   \end{figure*}

%
%
In Fig.~\ref{FIG:DUSE123} we present the evolution of quantities such as the total dust injection rates 
$E_{\mathrm{d,SN}}(t)$, $E_{\mathrm{d,AGB}}(t)$, the AGB and SNe rates  
$R_{\mathrm{AGB}}(t)$, $R_{\mathrm{SN}}(t)$, the SFR and the metallicity $Z(t)$ for a range of
initial galaxy gas masses. 
The rates and the metallicities are independent of the assumed dust formation efficiency and 
can be discussed for each initial mass $M_{\mathrm{ini}}$. 
The results for these quantities, which come from EIT08M, WW95M, and N06M with 
identical parameter setting, are very similar. 
Consequently these quantities are discussed based on EIT08M.
The upper row of Fig.~\ref{FIG:DUSE123} shows the dust injection rates of AGB stars $E_{\mathrm{d,AGB}}(t)$ 
and SNe $E_{\mathrm{d,SN}}(t)$. 
For supernovae $E_{\mathrm{d,SN}}(t)$ is highest for a Larson 2 IMF and lowest for a Salpeter IMF. 

The SN dust production rates are approximately one order of magnitude lower for 
$\epsilon_{\mathrm{low}}(m)$  than for $\epsilon_{\mathrm{high}}(m)$. 
 Similar difference is the case between  $\epsilon_{\mathrm{high}}(m)$ and  $\epsilon_{\mathrm{max}}(m)$.
In all cases a clear separation of the values of $E_{\mathrm{d,SN}}(t)$ amongst the various IMFs 
is  pronounced throughout the evolution.  

The dust injection rates for AGB stars, $E_{\mathrm{d,AGB}}(t)$, are considerably influenced 
by the long lifetimes. 
After $\sim$ 200 Myr the AGB dust production rates $M_{\mathrm{d,AGB}}(t)$ are comparable 
to the dust injection rates for SNe with $\epsilon_{\mathrm{high}}(m)$.
This is caused by the higher dust production efficiency for AGB stars between $\sim$ 3--4 $\Msun$,
leading to larger amounts of dust produced 
than for AGB stars in the mass range of 4--8 $\Msun$ (see Sect.~\ref{SSC:EDESNAGB}). 
The small variations in $E_{\mathrm{d,AGB}}(t)$ are mainly due to alteration of the stellar yields and 
dust formation efficiencies at different metallicities. 
 
The SNe and AGB dust injection rates decline as a consequence of decreasing 
SNe and AGB rates, as shown in Fig.~\ref{FIG:DUSE123} (second row). Both rates  exhibit a faster 
decline for the lower massive galaxies and for IMFs biased towards the intermediate and low mass stars. 

This behaviour is determined by the slope of the SFR, which depends on the initial mass 
of the galaxy (see Fig.~\ref{FIG:DUSE123} third row). 
The decline in the SFR for systems with 
$M_{\mathrm{ini}}$ $\ge$ 3 $\times$ $10^{11}$ $\Msun$ is shallower. 
For these galaxies, a high SFR of a few hundred $\Msun$  yr$^{-1}$ can be sustained over at least 1Gyr of evolution. 
For the lower mass galaxies, the SFR declines faster and exhibits a strong dependence on the IMFs.  
At a time of 400 Myr the SFR for a galaxy with $M_{\mathrm{ini}}$ = 1 $\times$  $10^{11}$ $\Msun$ 
is between 60 (Salpeter) to 500 (Larson 2) $\Msun$  yr$^{-1}$.
When $M_{\mathrm{ini}}$ = 5 $\times$  $10^{10}$ $\Msun$, $\psi(t)$ = 20--400 $\Msun$  yr$^{-1}$. 
After $\sim$ 1 Gyr the difference between the SFRs obtained with a Larson 2 IMF and 
a Salpeter IMF is more than an order of magnitude. 
   
In Fig.~\ref{FIG:DUSE123} (bottom row) it is seen that the metallicity $Z(t)$ in the two lower 
mass galaxies rises quickly within the first 100--200 Myr up to values of more than 
5 $\Zsun$ for either a top-heavy or a Larson 2 IMF. Thereafter it remains rather constant.
For the system with $M_{\mathrm{ini}}$ = 5 $\times$ $10^{11}$ $\Msun$, the metallicity 
increases more slowly than in the lower mass galaxies. 
At an epoch of 400 Myr, a metallicity of  2-3 $\Zsun$ can be observed 
for the top-heavy and Larson 2 IMFs,
while the Salpeter IMF achieves only a bit less than half solar. 
In the most massive galaxies, the metallicity exceeds $\Zsun$ only in case of the top-heavy IMFs 
before an age of 400--500 Myr, but remains below $\Zsun$ for the IMFs favouring lower mass stars.  
%
%
\subsection{Evolution of gas, metals and stellar masses}
\label{SSC:EGM}
\begin{figure}
   \centering
   \includegraphics[width=9 cm ]{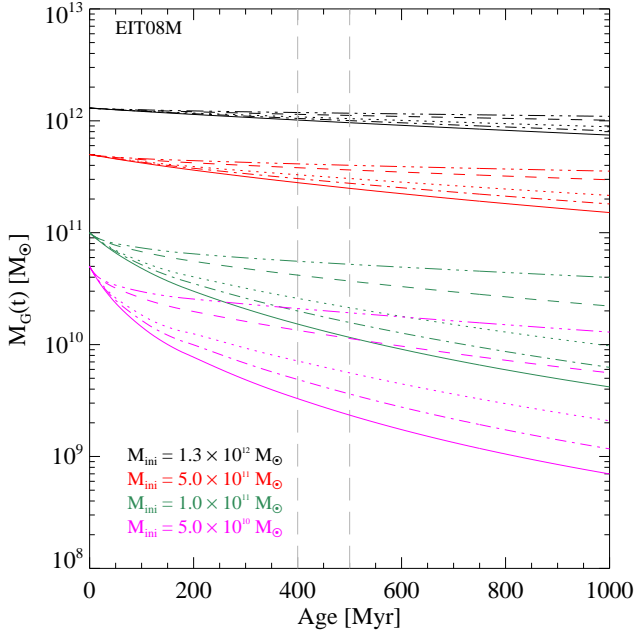}
      \caption{ Evolution of the gas mass $M_{\mathrm{G}}(t)$ based on EIT08M. 
      Results are shown for $M_{\mathrm{ini}}$ = 1.3 $\times$ 10$^{12}$ $\Msun$ (black lines), 
      5 $\times$ 10$^{11}$ $\Msun$ (red lines),
      1 $\times$ 10$^{11}$ $\Msun$ ( green lines), and 
      5 $\times$ 10$^{10}$ $\Msun$ (magenta lines). 
      The grey vertical dashed lines indicate epochs at  400 and 500 Myr after the onset of starburst.
      The solid, dotted, dashed, dashed-dotted, and dashed-dot-dotted lines represent 
      the Salpeter, mass-heavy, top-heavy, Larson 1, and Larson 2 IMF, respectively. 
              }
     \label{FIG:PLGASMAS}
   \end{figure}

%
%
The evolution of quantities, such as the gas mass, mass of metals, and  stellar masses 
is discussed based on EIT08M.
The Fig.~\ref{FIG:PLGASMAS} illustrates the evolution of the H + He gas mass $M_{\mathrm{G}}(t)$.  
As a consequence of the scaling of the SFR with the mass of the ISM, 
$M_{\mathrm{ISM}}(t) \equiv M_{\mathrm{G}}(t)$ $+$ $M_{\mathrm{Z}}(t)$, 
the progression of $M_{\mathrm{G}}(t)$ is identical to the SFR described 
in Sect. \ref{SSS:EDRSM}. 

The most massive galaxy with $M_{\mathrm{ini}}$ = 1.3 $\times$  $10^{12}$ $\Msun$ 
has a residual gas mass $M_{\mathrm{G}}(t)$ of $\sim 10^{12}$ $\Msun$ after 1 Gyr. 
This explains the very low dust-to-gas ratio for such a system 
even though very high dust masses can be reached. 
The galaxy with $M_{\mathrm{ini}}$ = 5 $\times$  $10^{11}$ $\Msun$ gets more exhausted, 
but also retains a gas mass $M_{\mathrm{G}}(t)$ of 3--4 $\times$  $10^{11}$ $\Msun$ 
after $\sim$ 400--500 Myr. 

In analogy with the curves of the SFR, the curves of the gas mass for the system with 
$M_{\mathrm{ini}}$ = 1 $\times$  $10^{11}$ $\Msun$ feature a 
pronounced separation between the IMFs. 
A Larson 2 IMF shows a flat evolution leading to a slower depletion of the gas than for the 
Salpeter IMF and after 400--500 Myr $\sim$ 5 $\times$ $10^{10}$ $\Msun$ of gas is still present. 
The lowest mass system ($M_{\mathrm{ini}}$ = 5 $\times$ $10^{10}$ $\Msun$) 
exhibits the strongest dependence on the IMF. 
After 400 Myr, $M_{\mathrm{G}}(t)$ is 
$\sim$ 2 $\times$ $10^{10}$ $\Msun$ for a Larson 2 IMF. 
The difference between a Salpeter IMF and a Larson 2 IMF after one Gyr is an order of magnitude. 
The residual gas mass with a Salpeter IMF is a few times $10^{8}$ $\Msun$ at this time. 

The steepness of the decline in the progression of $M_{\mathrm{G}}(t)$ 
is influenced by the stellar feedback.
For either a Larson 2 or a top-heavy IMF, more massive stars are formed. 
Such stars live short lives and release most of their mass 
back into the ISM, through either stellar winds or in explosive events. 
The IMFs favouring lower masses, however, lock most of the gas 
used for star formation into intermediate-to-low-mass stars. 
These stars are formed at higher rates than massive stars. 
The low-mass stars live a long time and also do not inject much material back  
into the ISM during our considered time span of 1 Gyr. 
As a consequence, the material available for star formation gets more 
rapidly depleted for the IMFs favouring low-mass stars 
than for the top-heavy IMFs. 
Hence, the evolution of the gas mass and the SFR results in a steeper decline. 

\begin{figure*}
   \centering
   \includegraphics[width=\textwidth]{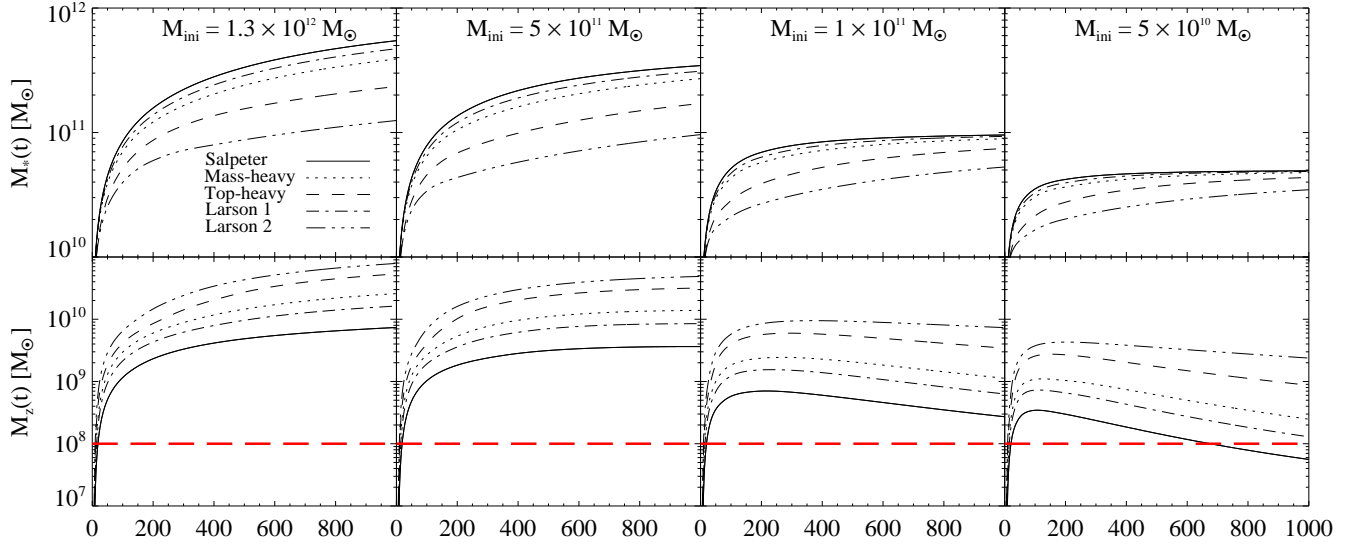}
      \caption{Evolution of the stellar mass and mass of metals based on EIT08M.
      Results are shown for $M_{\mathrm{ini}}$ = 1.3 $\times$ 10$^{12}$ $\Msun$ (first column), 
      5 $\times$ 10$^{11}$ $\Msun$ (second column),
      1 $\times$ 10$^{11}$ $\Msun$ (third column), and 
      5 $\times$ 10$^{10}$ $\Msun$ (fourth column). 
      Top row: Stellar mass $M_{\mathrm{\ast}}(t)$. 
       Bottom row: Mass of metals $M_{\mathrm{Z}}(t)$. 
      The red lines mark a metal mass of $10^{8}$ $\Msun$. 
      The solid, dotted, dashed, dashed-dotted, and dashed-dot-dotted lines represent 
      the Salpeter, mass-heavy, top-heavy, Larson 1, and Larson 2 IMF, respectively.  
              }
     \label{FIG:DUSE123_M}
   \end{figure*}

This contrasts with the stellar masses $M_{\mathrm{\ast}}(t)$ (Fig.~\ref{FIG:DUSE123_M} top row)
obtained from the relation $M_{\mathrm{\ast}}(t) = M_{\mathrm{ini}} - M_{\mathrm{ISM}}(t)$. 
For the IMFs favouring lower mass stars, $M_{\mathrm{\ast}}(t)$ rises steeply and approaches 
the initial mass of this galaxy in the case of the lower massive galaxy systems.  
The slope of $M_{\mathrm{\ast}}(t)$ for the top-heavy IMFs is shallower, and lower stellar masses 
$M_{\mathrm{\ast}}(t)$ are achieved.  

The Fig.~\ref{FIG:DUSE123_M} ( bottom row) depicts the mass of the ejected 
heavy elements $M_{\mathrm{Z}}(t)$. 
For systems with $M_{\mathrm{ini}}$ $>$ 1 $\times$  $10^{11}$ $\Msun$, 
the metal enrichment in the ISM is considerable. 
The mass of the heavy elements increases up to a few times $10^{10}$ $\Msun$.  

In the lower mass galaxies $M_{\mathrm{ini}}$ $\le$ 1 $\times$  $10^{11}$ $\Msun$, 
the amount of metals $M_{\mathrm{Z}}(t)$ attained 
reaches a maximum within the first 200 Myr whereafter $M_{\mathrm{Z}}(t)$ declines. 
This is caused by astration. 
The maximum mass of metals obtained with a Salpeter IMF is only a few times $10^{8}$ $\Msun$. 
For the lowest mass galaxy the amount of metals after 
$\sim$ 600 Myr for a Salpeter IMF is less than $10^{8}$ $\Msun$. 
This implies that dust masses $M_{\mathrm{d}}(t)$ in excess of $10^{8}$ $\Msun$ 
for this IMF are unfeasible, even if dust-grain growth in the ISM is invoked. 

%
%
 %
\subsection{Comparison to other dust evolution models for high-$z$ galaxies}
\label{SEC:COMP}
%
%
The models with a `maximum' SN dust production efficiency for a system with 
$M_{\mathrm{ini}}$ of 5 $\times$ $10^{10}$ $\Msun$ (Fig.~\ref{FIG:DUSE15} two left columns, first row) 
can be compared to models of \citet{dwe07}. 
For this efficiency the AGB dust production is negligible and the SN dust yields are similar to the dust yields used by  \citet{dwe07}. 
Our results for a top-heavy IMF and $M_{\mathrm{cl}}$ = 0 at an epoch of 400 Myr 
agrees with their results, 
while they disagree for a Salpeter IMF and for cases with $M_{\mathrm{cl}}$ = 100 $\Msun$.  

The origin of this disagreement can be traced to the neglect of the lifetime-dependent mass injection of stars 
and a different treatment of the mass recycled into the ISM in the models of \citet{dwe07}.
The latter is approximated as 0.5$\psi(t)$ in their models, which leads to an IMF-independent evolution of the ISM mass. 
In the case of a Salpeter IMF, this simple approximation has consequences for the evolution of all physical properties. 
It implicitly presupposes that even stars, which are formed between the lower mass limit of the IMF and 
the lower SN mass limit, immediately return half of their stellar mass back into the ISM.  
However, stars between 3--8 $\Msun$ eject their elements up to a few 100 Myr delayed and
stars $\lesssim$ 3 $\Msun$ do not eject a significant amount of elements within the first Gyr. 
For a Salpeter IMF, where more low-mass stars than high-mass stars are formed, we have shown  
that the mass of the ISM gets exhausted more rapidly than for the IMFs favouring higher masses (see Sect.~\ref{SSC:EGM}).  
This results in a steeper decline in the SFR, the SN rates and the SN dust injection rates. 
Thus, our models lead to lower dust masses than the models by  \citet{dwe07}
for a Salpeter IMF. 

The model for a system with $M_{\mathrm{ini}}$ = 1.3 $\times$ 10$^{12}$ $\Msun$ and  
$M_{\mathrm{cl}}$  = 800 $\Msun$ (Fig.~\ref{FIG:WW95CONST} top left), a Larson 1 IMF and 
the `high' SN efficiency is directly comparable to \citet{val09}. 
Our results are in good agreement with their results, 
although our model does not quite reach $10^8$ $\Msun$. 
This discrepancy may stem from differences in the SN dust yields used.
Additionally, the treatment of stars between 40--100 $\Msun$ by  \citet{val09} is not unambiguously traceable.   
\subsection{Results for models including a SMBH}
\label{SEC:MSMBH} 
%
%
%
We have investigated whether including the formation of the SMBH leads to
differences in the evolution of either the total amount of dust or the physical properties of a galaxy.
Based on the observed SMBH masses $\ge$ $10^{9}$ $\Msun$  for high-redshift QSOs \citep[e.g.,][]{will03, vest04, jiang06, wan10}, 
we studied two cases for the SMBH mass,  $M_{\mathrm{SMBH}}$ = 3 $\times$ $10^{9}$ $\Msun$ 
and $M_{\mathrm{SMBH}}$ = 5 $\times$ $10^{9}$ $\Msun$.
According to \citet{kaw09}, the final SMBH mass takes up approximately 1--10 \% of 
the supply mass $M_{\mathrm{SMBHsup}}$.
This implies that to grow a SMBH of 3(5) $\times$ $10^{9}$ $\Msun$ the minimum 
required supply mass ranges between 
3(5) $\times$ $10^{10}$ to 3(5) $\times$ $10^{11}$ $\Msun$. 
In view of our assumption that $M_{\mathrm{SMBHsup}}$ $\equiv$ $M_{\mathrm{ini}}$,  we included 
the SMBH formation in galaxies with $M_{\mathrm{ini}}$ = 5 $\times$ 10$^{10}$ $\Msun$ and $M_{\mathrm{ini}}$ = 1 $\times$ 10$^{11}$ $\Msun$. 

\begin{figure}
   \centering
   \resizebox{\hsize}{!}{ \includegraphics{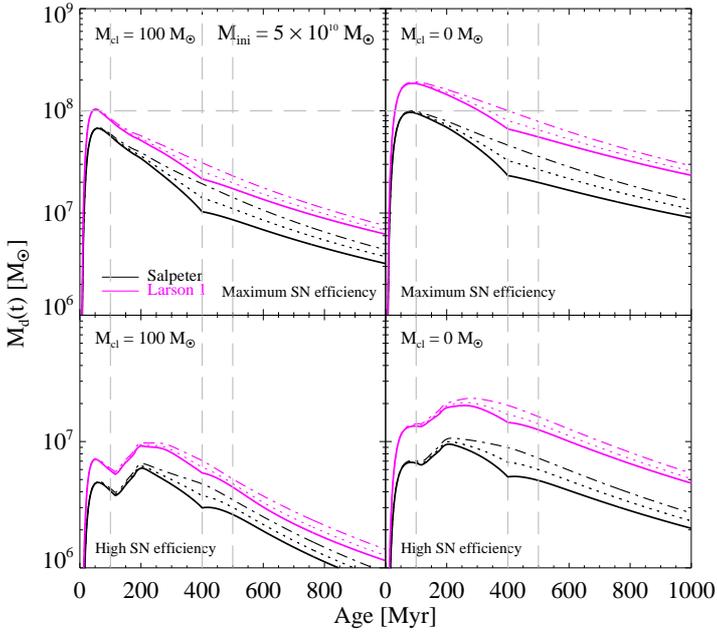}}    
      \caption{Comparison of  the evolution of the dust mass between EIT08M 
       including and without the SMBH formation.
      The initial gas mass of the galaxy $M_{\mathrm{ini}}$ = 5 $\times$ 10$^{10}$ $\Msun$. 
      Calculations are shown for a `maximum' (top row) and
      a `high' (bottom row) SN dust production efficiency $\epsilon_{\mathrm{i}}(m)$. 
      Dust destruction is taken into account for   
      $M_{\mathrm{cl}}$ = 100 $\Msun$ (left column) and 0 $\Msun$ (right column). 
       Calculations including the SMBH are performed for 
       a SMBH mass $M_{\mathrm{SMBH}}$ = 5 $\times$ $10^{9}$ $\Msun$ (solid curves) and
       for $M_{\mathrm{SMBH}}$ = 3 $\times$ $10^{9}$ $\Msun$  (dotted curves). 
      The calculations without the SMBH are illustrated as dot-dashed curves. 
      The black lines represent the Salpeter IMF and the magenta lines the Larson 2 IMF.
       The grey horizontal dashed line marks the limit of $10^8$ $\Msun$ of dust. 
      The grey vertical dashed lines indicate epochs at 100, 
      400, and 500 Myr after the onset of starburst. 
       }
     \label{FIG:PLSMBHSFR3E}
   \end{figure}

In Fig.~\ref{FIG:PLSMBHSFR3E} we show the results of 
$M_{\mathrm{d}}(t)$ for a Salpeter and Larson 1 IMF 
for an EIT08M and  $M_{\mathrm{ini}}$ = 5 $\times$ $10^{10}$ $\Msun$ 
where the SMBH formation has been included. 
These results are compared to those of a model with the same $M_{\mathrm{ini}}$, 
but without a SMBH. 

Taking the SMBH into account leads to a steeper decline in $M_{\mathrm{d}}(t)$ and therefore to a lower 
amount of dust than for the model without a SMBH after $\sim$ 100 Myr.
All models are similar within the first $\sim$ 100 Myr. 
For $M_{\mathrm{SMBH}}$ = 5 $\times$ $10^{9}$ $\Msun$ the difference 
is at most $\sim$ 50 \% for a Salpeter IMF and $\sim$ 30 \% for a Larson 1 IMF
at an epoch of 400 Myr.  
At the same epoch but for $M_{\mathrm{SMBH}}$ = 3 $\times$ $10^{9}$ $\Msun$, the dust mass for a Salpeter IMF is reduced by only about 30 \%, and 20 \% for the Larson 1 IMF. 
In cases of IMFs biased towards higher masses and in the more massive galaxy systems, the formation of the 
SMBH, as introduced here,  
does not noticeably effect the progression of $M_{\mathrm{d}}(t)$ and the physical properties of a galaxy.   
%
 %
%
\section{Discussion}
\label{SEC:DISC}
\begin{figure*}
   \centering
   \includegraphics[width=\textwidth]{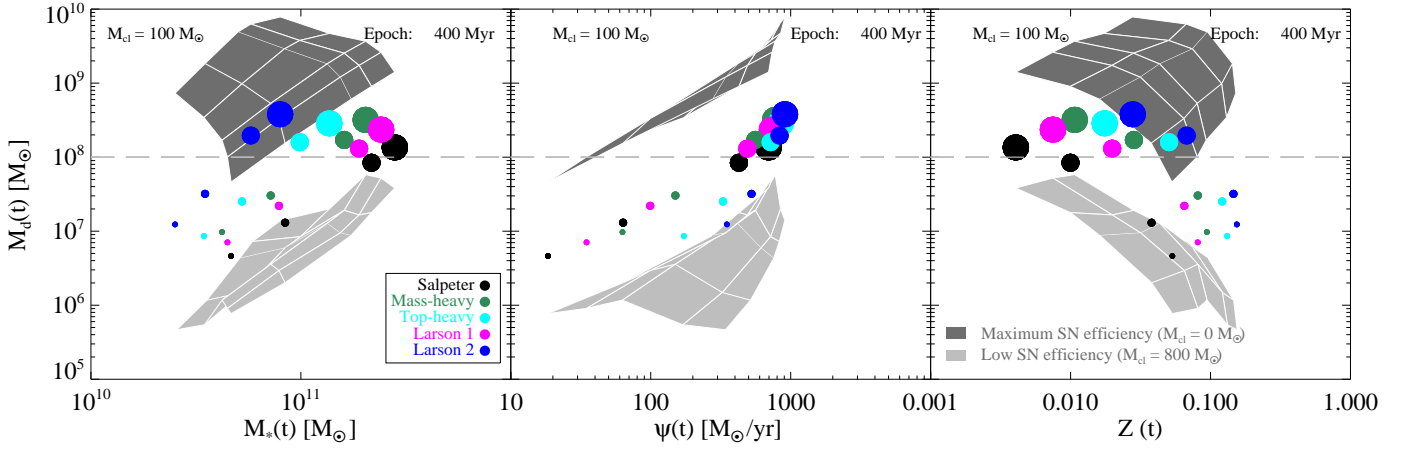}      
      \caption{Relation between dust mass, stellar mass, SFR, and metallicity in correlation with 
      the total mass of the galaxy and the IMF for an epoch at 400 Myr. 
      Left panel: Dust mass $M_{\mathrm{d}}(t)$ versus stellar mass $M_{\mathrm{\ast}}(t)$. 
      Middle panel: Dust mass $M_{\mathrm{d}}(t)$ versus SFR $\psi(t)$. 
      Right panel: Dust mass $M_{\mathrm{d}}(t)$ versus metallicity $Z(t)$.
      The dark grey shaded area marks the dust masses $M_{\mathrm{d}}(t)$ obtained with a `maximum' 
      SN dust production efficiency  $\epsilon_{\mathrm{max}}(m)$ and no dust destruction 
      in the ISM ($M_{\mathrm{cl}}$ = 0 $\Msun$). 
      The light grey shaded area marks dust masses $M_{\mathrm{d}}(t)$ for the `low' 
      SN dust production efficiency $\epsilon_{\mathrm{low}}(m)$ 
      with dust destruction in the ISM ($M_{\mathrm{cl}}$ = 800 $\Msun$). 
      Cross points of the thin white lines indicate the data points for each IMF and mass 
      $M_{\mathrm{ini}}$ of the galaxy.
      The coloured filled circles signify the dust masses obtained for a `high' SN efficiency  
      $\epsilon_{\mathrm{high}}(m)$ and a dust destruction in the 
      ISM  with $M_{\mathrm{cl}}$ = 100 $\Msun$. The size of the circles is scaled by the initial mass 
      $M_{\mathrm{ini}}$ of the galaxy. 
      The black, green, cyan, magenta, and blue colors denote the Salpeter, mass-heavy, top-heavy, 
      Larson 1, and Larson 2 IMF, respectively. 
                             }
     \label{FIG:STARDUST}
   \end{figure*}

%
 %
 %
We have shown that, depending on the assumptions for certain model parameters, dust masses 
in excess of $10^{8}$ $\Msun$ can be obtained in our model. 
All models leading to dust masses $\ge$ $10^{8}$ $\Msun$ within the first 400 Myr are listed in Table~\ref{TAB:FIT}.
In Fig. \ref{FIG:STARDUST} we show the resulting relations between the dust mass and 
stellar mass, SFR, and metallicity at 400 Myr.

It is evident that the dust yields are related to the mass of the galaxy. 
For a given combination of  IMF, SN dust production efficiency and $M_{\mathrm{cl}}$ higher dust yields 
are obtained with increasing $M_{\mathrm{ini}}$.
In the discussion below we consider
$\epsilon_{\mathrm{high}}$ with $M_{\mathrm{cl}}$ = 100 $\Msun$ 
as the reference case. 

The left panel in Fig. \ref{FIG:STARDUST} depicts the dust mass versus the stellar mass. 
For a given IMF the stellar masses increase with $M_{\mathrm{ini}}$. 
However for a given $M_{\mathrm{ini}}$, a wide spread of the stellar mass is found 
with varying IMF. 
We find that for the reference case, galaxies with $M_{\mathrm{ini}}$ $\gtrsim$ 5 $\times$  $10^{11}$ $\Msun$ and  
an IMF biased towards higher masses reproduce dust masses $>$ $10^{8}$ $\Msun$, 
while this is only possible in the most massive system with a Salpeter IMF. 
The stellar masses for these galaxies and with the more top-heavy IMFs are in good agreement 
with observations of galaxies at high-$z$.    
\citet{sant10} and \citet{mich10a} 
shows that, out of a sample of high-$z$ submillimetre galaxies and local ULIRGs, most of 
these are assembled at stellar masses around $10^{11}$ $\Msun$ and have dust masses of $10^{8-10}$ $\Msun$. 

Alternatively, a `maximum' SN dust production efficiency can account for the required dust mass. 
For this SN efficiency, even a higher dust destruction in the ISM of $M_{\mathrm{cl}}$ = 800 $\Msun$ can be accomodated. 
However, assumptions of either $\epsilon_{\mathrm{max}}$ or $M_{\mathrm{cl}}$ = 800 $\Msun$ are both controversial. 
So far only in the SN remnants Cas A \citep{wils05,dun09} and Kepler \citep{gom09} have dust masses been claimed 
that are consistent with `maximum' SN efficiency, $\epsilon_{\mathrm{max}}$. 
Most SN dust observations reveal dust masses implying efficiencies in the range of 
$\epsilon_{\mathrm{low}}$--$\epsilon_{\mathrm{high}}$ (see GAH11).
Additionally, theoretical models  \citep{noz10} predict that depending on the density and geometry of the CSM and shocks, 
only a part of the dust may survive the destructive reverse shock. 

The uncertainty of dust destruction through SN shocks has been discussed in Sect.~\ref{SSC:DISM}. 
The results obtained in this work strengthen the concerns of a high value for $M_{\mathrm{cl}}$. 
Apart from the `maximum' SN efficiency, for a high value of $M_{\mathrm{cl}}$ = 800 $\Msun$, 
only the combination of a `high' SN efficiency $\epsilon_{\mathrm{high}}$ with  
$M_{\mathrm{ini}}$ = 1.3 $\times$  $10^{12}$ $\Msun$ leads to dust masses close to $10^{8}$ $\Msun$, 
regardless of the IMF.  
However the remaining gas mass $M_{\mathrm{G}}(t)$ is $>$ $10^{12}$ $\Msun$, 
implying dust-to-gas mass ratios $\eta_{\mathrm{d}}(t) < 10^{-4}$, is not supported by observations. 
The low dust-to-gass mass ratio is also, why dust destruction in the ISM is less efficient than 
in the lower mass galaxy systems. 

In principle, the fairly sensitive interplay between dust-to-gas ratio $\eta_{\mathrm{d}}(t)$ and SN rates 
might be important when contemplating additional dust source in less massive galaxies. 
 As long as SN rates are high, any increase in the dust mass, and hence in $\eta_{\mathrm{d}}(t)$, 
 leads to higher destruction rates.  
This shows that for $M_{\mathrm{cl}}$ = 800 $\Msun$ it is difficult to reach high amounts of dust, 
unless a rapid enrichment with high dust masses from either stellar sources or grain growth in the ISM takes place. 
In the latter case, grain growth rates must be comparable to SN dust injection rates for $\epsilon_{\mathrm{max}}$.   
Alternatively, a lowering of the dust-to-gas ratio $\eta_{\mathrm{d}}(t)$ due to, for example, infalling gas might be an option.
However this alternative results in an increased total mass of the galaxy anyway.

Given these uncertainties, we find that the reference case constitutes the most likely scenario. 
In Fig.~\ref{FIG:STARDUST} (middle panel) we have plotted the dust mass versus SFR. 
The highest SFRs can be sustained for the high-mass weighted IMFs in the more massive galaxies. 
For the less massive systems, also with top-heavy IMFs, the SFR declines rather fast.
In comparison to observations of galaxies  containing high dust masses at $z$ $>$ 6, we find that all 
systems more massive than 1 $\times$  $10^{11}$ $\Msun$ can be brought into agreement with the SFRs  
derived from observations \citep[e.g.,][]{bertol03, dwe07, wan10}. 
The SFRs inferred from the observed far-infrared luminosities are naturally uncertain 
and sensitively dependent on assumptions about the IMF, the duration of the starburst, 
and the contribution of the active galactic nucleus (AGN) \citep[e.g.,][]{omo01}. 
Commonly a  Salpeter IMF is used, leading to high SFRs up to  a few times 10$^{3}$ $\Msun$ yr$^{-1}$ \citep[e.g.,][]{walt04, wan10}. 
As pointed out by \citet{dwe07}, the inferred SFR however decreases to about 400 $\Msun$ yr$^{-1}$ if a top-heavy IMF is assumed. 
In Fig.~\ref{FIG:STARDUST} (middle panel) it is evident that galaxies between 1--5 $\times$  $10^{11}$ $\Msun$, and top-heavy IMFs result in SFRs between about 300--700 $\Msun$ yr$^{-1}$.

The right panel of Fig.~\ref{FIG:STARDUST} shows that the metallicity decreases
with increasing dust mass.
IMFs favouring low masses lead to lower metallicities than the more top-heavy IMFs. 
The IMFs favouring low masses lock most material used for star formation in low-mass stars. 
These stars, however, do not recycle material back into the ISM, which therefore gets rapidly exhausted. 
In turn this leads to a high stellar mass $M_{\mathrm{\ast}}(t)$.  
The more top-heavy IMFs form more short lived massive stars at higher rates. 
These stars recycle a copious amount of their mass back into the ISM. 
Therefore these IMFs lead to lower stellar masses $M_{\mathrm{\ast}}(t)$, since the remnant mass of SNe is low.  
In addition, massive stars enrich the ISM with metals, while owing to their longer lifetimes, low-mass stars do not eject 
a significant amount, if any at all, of heavy elements back into the ISM  within 1 Gyr.  
This leads to lower metallicity for the lower mass star weighted IMFs, and higher metallicity for the top-heavy IMFs.

Metallicities $>$ $\Zsun$ have been found in strong star-forming galaxies, such as ULIRGs or submillimetre galaxies, 
as well as in high-$z$ QSOs \citep[e.g.,][]{fan03, freu03, kawa10}. 
In comparison to our calculated models, such metallicities can only be reached with the IMFs biased towards higher masses. 
In the lowest mass systems with these IMFs, the metallicity reaches values of $\sim$ 5 $\times$ $\Zsun$. 
\begin{table}[!bt]
\caption{{EIT08M exceeding $10^{8}$ $\Msun$ of dust within $\sim$ 400--500 Myr}}          
\label{TAB:FIT}
\centering
\begin{tabular}{llrl}
\hline
\hline
    $M_{\mathrm{ini}}$
   & $\epsilon_{\mathrm{SN}}$
   & $M_{\mathrm{cl}}$
   & IMF \\                 
\hline
\\1.3 $\times$ $10^{12}$ $\Msun$	&max	&0		&all\\
							&max	&100	&all\\
							&max	& 800	&all\\
							&high	&0		&all\\
							&high	&100	&all\\
							&low		&0		&mass-heavy, Larson 1\\
							&low		&100	&mass-heavy, Larson 1\\		

  5 $\times$ $10^{11}$ $\Msun$	&max	&0		&all\\
  							&max	&100	&all\\
							&max	& 800	&all\\
							&high	&0		&all\\
							&high	&100	&top-heavy, Larson 1, 2, mass-heavy\\
							&low		&0		&mass-heavy\\
 							
  1 $\times$ $10^{11}$ $\Msun$	&max	&0		&all\\
  							&max	&100	&top-heavy, Larson 1, 2, mass-heavy\\
							&high	&0		& Larson 2\\

  5 $\times$ $10^{10}$ $\Msun$	&max	&0		&top-heavy, Larson 1, 2, mass-heavy\\
  							&max	&100	& Larson 2\\
							&max\tablefootmark{a}	&100		&top-heavy, Larson 1, mass-heavy\\
\hline
\end{tabular}
\tablefoot{
\tablefoottext{a}{Top-heavy IMF: only within the first $\sim$ 300 Myr, Larson 1 and mass-heavy IMF only within the first $\sim$ 100 Myr}
}
\end{table}
 %
%
%
\subsection{Caveats in our approach} 
\label{SEC:NSEC} 
%
The presented models are based on rather simple assumptions, such as a closed box environment, a common
constant initial SFR of the starburst, and a very simple treatment of the SMBH growth. 
With these assumptions the models are comparable to similar works \citep[e.g.,][]{dwe07, val09}, and  
diverse evolutionary trends could be investigated in more detail, as discussed in previous sections. 

However, such models usually do not capture possible impacts on the evolution of the galaxy 
arising from galaxy mergers or gas flows powered by, for example, SNe or the SMBH. 
Different evolutionary paths likely caused by these effects and the mass of the galaxy possibly 
result in starburst intensities different from our assumption, in turn leading to a different temporal progression of various quantities. 
Furthermore, the growth of the SMBH is linked to several physical processes, e.g., the energy feedback of SNe, 
and realistically may not take place at a constant growth rate.  
Despite the neglect of these effects in our model, some general remarks about their influence can be made based on our results.  

For example, infall rates that could affect the systems might have to be unrealistically high because of the very high SFRs. 
However, merging galaxies might have an effect. 
The same applies to outflows in the host galaxy or the feedback from the SMBH into the ISM.
In our models the loss of material to fuel the SMBH can in principle be interpreted as some constant `outflow' of the host galaxy.  
Although we introduced a rough treatment for the SMBH growth, we could show that `outflow'  rates of the host galaxy of 
about 7.5--12.5 $\Msun$ yr$^{-1}$ on a timescale of 400 Myr only affect the amount of dust in the least massive systems 
(and IMFs biased towards low-mass stars), but are negligible in the larger galaxies.  
In this regard, it remains to be investigated whether the energy deposition by SNe/AGNs in the ISM 
could be high enough to initiate sufficiently higher outflow rates to impact the total amount of dust in a galaxy. 

A change in the evolution of the dust mass and other properties may also result from quasar winds, 
in which dust formation has also been suggested \citep{elv02}.
According to \citet{elv02}, the estimated mass-loss rates of about $>$ 10 $\Msun$ yr$^{-1}$ in  
the most luminous quasars with luminosities $>$ 10$^{47}$ ergs s$^{-1}$  
\citep[e.g.,] [] {omo01, omo03, caril01, bertol02} imply an amount of $\sim$ 10$^{7}$ $\Msun$ of 
dust produced over 10$^{8}$ yr.  
Comparing these rates to our model results, we find that neither the mass-loss rates nor the dust mass 
seem to be high enough to be relevant for the evolution of dust and other properties.   

We find that a strong impact on the evolution of dust is caused by the dust sources. 
Despite the included detailed treatment of the dust contribution from stellar sources in the developed model,   
the poorly understood dust production by SNe, but also dust destruction by SN shock interactions, 
very likely constitute the largest uncertainties in the evolution of dust. 
Although not included in the model, alternative dust sources such as dust grain growth in the ISM might be relevant \citep[e.g.,][]{dwe07, drai09, mich10b}.
%
%
\section{Conclusions}
\label{SEC:CONCL}
%
%
In this work we have developed a chemical evolution model for starburst galaxies at high redshift. 
The main purpose is to investigate the evolution of the dust content arising from SNe and AGB stars 
on timescales less than 1 Gyr. 
In addition, we elaborated on the evolution of several physical properties of galaxies. 
The model allows the exploration of a wide range of parameters. 
The main parameters varied in this work are the mass of the galaxy, the IMF, 
SN dust production efficiencies, and dust destruction in the ISM through SN shocks and stellar yields. 
The main results can be summarized as follows.
\begin{enumerate}
\item The total amount of dust and the physical properties of a galaxy are strongly dependent on the IMF 
and correlate with the mass of the galaxy. 
For many properties we find an increasing disparity between the IMFs with decreasing mass of the galaxy. 
Higher dust masses are obtained with increasing mass of the galaxy. 

\item The maximum dust masses can be obtained with IMFs biased towards higher mass stars and a 
`maximum' SN dust production efficiency in galaxies with masses between 5 $\times$  $10^{10}$ $\Msun$ 
and 1.3 $\times$  $10^{12}$ $\Msun$. 
This is found independently of the strength of dust destruction in the ISM.  
The case of `maximum' SN efficiency and no destruction constitute the maximum possible dust masses 
attainable with SNe and AGB dust production.  
The case with `low' SN efficiency and a high dust destruction (M$_{\mathrm{cl}}$ = 800 $\Msun$) 
gives the lowest possible dust masses. 

 \item The contribution by AGB stars is most visible in cases where a `low' SN dust formation efficiency is considered. 
 In this case the mass-heavy IMF dominates in all considered initial masses M$_{\mathrm{ini}}$, 
 and leads to the highest dust masses. 
 However, dust production with this efficiency is found to be insufficient to fully account for 
 dust masses in excess of $10^{8}$ $\Msun$ within 400 Myr. 
 In the galaxies with $M_{\mathrm{ini}}$ $\gtrsim$  5 $\times$  $10^{11}$ $\Msun$, this limit can be reached 
 at epochs $\gtrsim$ 400 Myr if there is no dust destruction in the ISM.

\item  The total dust mass is considerably reduced when destruction by SN shock waves 
is taken into account. The strength of destruction  in the ISM is given in terms of the mass $M_{\mathrm{cl}}$, 
which is the mass of the ISM that is swept up and cleared of the containing dust through one SN remnant.   
Top-heavy IMFs are sensitive to dust destruction, 
while IMFs favouring lower masses are more resistent to destruction due to lower SN rates.
This leads to a significantly greater reduction and, in some cases, to a lower amount of  dust 
with the top-heavy IMFs. 
       
\item  
At early epochs ($<$ 200 Myr)  SNe are primarily responsible for a significant enrichment by dust.
For a `high' SN dust production efficiency SNe can generate $10^{8}$ $\Msun$  within the first 100 Myr. 
In comparison, for `low' SN dust production efficiency, SNe are only able to increase 
the dust mass up to a few times $10^{6}$ $\Msun$ in the first 100--150 Myr.

\item Taking the growth of the SMBH into account leads to a reduction of the amount of dust of  at most $\sim$ 50 \%.
This is achieved for a SMBH mass $M_{\mathrm{SMBH}}$ = 5 $\times$ $10^{9}$ $\Msun$ in a galaxy with  
$M_{\mathrm{ini}}$  = 5 $\times$  $10^{10}$ $\Msun$ and only for IMFs favouring low-mass stars. 
In more massive systems and for top-heavy IMFs no variation in the dust mass and properties of the galaxy is encountered.

\item To account for dust masses $>$ $10^{8}$ $\Msun$ we find that galaxies with an initial gas mass of 
$M_{\mathrm{ini}}$  = 1--5 $\times$  $10^{11}$ $\Msun$ in connection with top-heavy IMFs, 
`high' SN dust production efficiency $\epsilon_{\mathrm{high}}$, and dust destruction 
in the ISM of  M$_{\mathrm{cl}}$ = 100 $\Msun$ are favoured. 
Models with the highest amount of $M_{\mathrm{ini}}$  = 1.3 $\times$  $10^{12}$ $\Msun$ and 
lowest of $M_{\mathrm{ini}}$  = 5 $\times$  $10^{10}$ $\Msun$ are discouraged. 

\end{enumerate}
More refined models including diverse flow scenarios tracing self-consistent SF histories, detailed SMBH growth scenarios,
and several plausible dust sources  are major future developments. 
%
%
 \begin{acknowledgements}
 
We would like to thank John Eldridge for providing tabulated values of his stellar evolution models. 
We also thank Justyn R. Maund,  Darach Watson, and Marianne Vestergaard for informative discussions. 
The Dark Cosmology Centre is funded by the DNRF.
            
\end{acknowledgements}
%
%
\bibliographystyle{aa}
\bibliography{reflist_ch_abb}
%
%
\end{document}